\documentclass[aps,prb,twocolumn,flotfix,showpacs,tightenlines,superscriptaddress]{revtex4-1}

\usepackage{amssymb}
\usepackage{graphicx}
\usepackage{braket}
\usepackage{amsmath}
\usepackage{bm}

\begin{document}
\title{Non-Markovian dynamics of a nanomechanical resonator measured by a
quantum point contact}
\author{Po-Wen Chen}
\author{Chung-Chin Jian}
\author{Hsi-Sheng Goan}
\email[Corresponding author: ]{goan@phys.ntu.edu.tw}
\affiliation{Department of Physics and Center for Theoretical Sciences, National Taiwan
University, Taipei 10617, Taiwan} 
\affiliation{Center for Quantum Science and Engineering, National Taiwan University,
Taipei 10617, Taiwan}


\begin{abstract}
We study the dynamics of a nanomechanical resonator (NMR) subject to a
measurement by a low-transparency quantum point contact (QPC) or tunnel
junction in the non-Markovian domain. We derive the non-Markovian
number-resolved (conditional) and unconditional master equations valid to
second order in the tunneling Hamiltonian
without making the rotating-wave approximation and the Markovian approximation,
generally made for systems in quantum optics. Our non-Markovian master equation
reduces, in appropriate limits, to various Markovian versions of master
equations in the literature. We find considerable difference in dynamics
between the non-Markovian cases and its Markovian counterparts. We also
calculate the time-dependent transport current through the QPC which
contains information about the measured NMR system. 
We find an extra transient current term proportional to the expectation 
value of the symmetrized product of the position and momentum operators of 
the NMR. This extra current term, with a coefficient coming from the 
combination of the imaginary parts of the QPC reservoir correlation 
functions, has a substantial contribution to the total transient current in 
the non-Markovian case, 
but was generally ignored in the 
studies of the same problem in the literature. Considering the contribution 
of this extra term, we show that a significantly qualitative 
and quantitative difference in the total transient current between the 
non-Markovian and the Markovian wide-band-limit cases can be
observed. Thus, it may 
serve as a witness or signature of the non-Markovian features in the coupled 
NMR-QPC system. 
\end{abstract}

\maketitle


\section{INTRODUCTION}

Recent advances in nanotechnology have enabled the fabrication of very small
quantum electronic devices that incorporate mechanical degrees of freedom,
called nanomechanical systems \cite{Graighead00,Roukes01,Blencowe04,Santamore04,Hensinger05,Wei06,Sun06}. The
interplay between electronic and mechanical degrees of freedom
has generated interesting dynamical effects \cite%
{Gorelik98,Park00,Blencoew2000,Utami04,Ouyang09}. These advances
have also opened a
new avenue to technology of high precision displacement measurement using
electronic devices, such as quantum dots, single electron transistors
(SET's), or quantum point contacts (QPC's) \cite{Blencoew2000,Mozyrsky
2004,Rodrigues 2005,Knobel 2003,LaHaye 2004,Schwab06,Mozyrsky 2002,Wabnig
2005,Clerk 2004,Smirnov 2003,Wabnig 2007,Rugar08}. Experiments using SET's
and QPC's have demonstrated displacement detections of a nanomechanical
resonator (NMR) with sensitivities close to the standard quantum limit \cite{Knobel 2003,LaHaye 2004,Schwab06,Rugar08}. Similar problems of a two-level
system measured by QPC's or SET's have also attracted much attention \cite{Gurvitz 1997,Shnirman98,Krotkov,Goan01,Goan03,Stace2004,Li 2004,Goan04,Li
2005,Ming2007,Rimberg03} theoretically and experimentally.

The transport properties in the nanostructure electronic devices are
often studied 
theoretically in the wide-band limit (WBL) and under the Markovian
approximation \cite{Blencoew2000,Mozyrsky 2004,Rodrigues 2005,Mozyrsky
2002,Wabnig 2005,Clerk 2004,Smirnov 2003,Wabnig 2007,Gurvitz
1997,Shnirman98,Krotkov,Goan01,Goan03,Stace2004,Li 2004,Goan04,Li 2005}. The
WBL approximation neglects an important fact that electron tunneling
amplitudes and also electrodes' densities of states are in general 
energy-dependent.
The Markovian approximation assumes that the
correlation time of the electrons in the electrodes (reservoirs) is much
shorter than the typical system response time. These approximations
may not be always true 
in realistic nanostructure devices. 
Hence, a recent development in quantum nanostructure electronic transport
has been devoted to the study of the non-Markovian effects on the electron
transport properties, taking into account the energy-dependent spectral
density of electrodes \cite{Ming2007,Taranko02,Guo05,Guo06,Guo08,Welack
  2006,Hou06,Jin10,Matisse 2008,Kleinekathofer2004,Zedler09}.
In this paper, we investigate the dynamics of a NMR subject to a
measurement by a low-transparency QPC or a tunnel junction in the
non-Markovian domain. This problem has
been extensively studied in the literature under various conditions and
approximations \cite{Mozyrsky 2002,Wabnig 2005,Clerk 2004,Smirnov
2003,Wabnig 2007}. 
In Ref.~\onlinecite{Mozyrsky 2002} a master equation of the reduced
density matrix 
of a NMR was obtained for zero-temperature QPC reservoirs (electrodes) in
the high-bias limit. 
The master equation presented 
in Ref.~\onlinecite{Clerk 2004} included not only the effect of the
QPC reservoirs in the high-bias limit 
but also the effect of a high-temperature thermal
environment on the NMR.
The master equation derived in Ref.~\onlinecite{Wabnig 2005} was claimed to be
applicable for a broad range of QPC temperatures and bias voltages and
for arbitrary NMR frequencies. However, the results
presented in these papers \cite{Mozyrsky 2002,Wabnig 2005,Clerk 2004,Smirnov
2003,Wabnig 2007} were under the Markovian approximation and without considering
the energy-dependent spectral density of electrodes. In this paper, we
take these into 
account and derive a time-local (time-convolutionless) non-Markovian master
equation \cite{Breuer2002,Paz
  2001,Breuer99,Schroder06,Shibata77,Kleinekathofer2004,Liu07,Sinayskiy09,Ferraro09,Mogilevtsev09,Haikka10,Ali10}
that reduces, in appropriate limits, to various Markovian versions 
of the master equations in these papers \cite{Mozyrsky 2002,Wabnig
  2005,Clerk 2004,Smirnov 2003,Wabnig 2007}. We find considerable
differences in dynamics between the 
non-Markovian case and its Markovian counterpart in some parameter regime. We
also calculate the time-dependent transport current through the QPC which
contains information about the measured NMR system. 
We find an extra transient current term proportional to the expectation 
value of the symmetrized product of the position and momentum operators of 
the NMR. This extra current term, with a coefficient coming from the 
combination of the imaginary parts of the QPC reservoir correlation 
functions, has a substantial contribution to the total transient current in 
the non-Markovian case and differs qualitatively and quantitatively from its 
Markovian WBL counterpart. But it was generally ignored in the 
studies of the same problem in the literature \cite{Mozyrsky
2002,Wabnig 2005,Clerk 2004,Smirnov 2003,Wabnig 2007}. 
Considering the contribution 
of this extra term, we show in this paper that a significantly qualitative 
and quantitative difference in the total transient current between the 
non-Markovian and the Markovian WBL cases can be observed. Thus, it may 
serve as a witness or signature of the non-Markovian features in the coupled 
NMR-QPC system.

The paper is organized as follows. In Sec.~\ref{sec:model}, we describe our
NMR-QPC model.  
In Sec.~\ref{n-resolvedME}, we
derive a time-local (time-convolutionless) non-Markovian
\cite{Breuer2002,Paz 2001,Breuer99,Schroder06,Shibata77,Kleinekathofer2004,Liu07,Sinayskiy09,Ferraro09,Mogilevtsev09,Haikka10,Ali10}
number-resolved or {\em n-} resolved (conditional) 
\cite{Gurvitz 1997,Shnirman98,Krotkov,Goan03,Li 2004,Goan04,Clerk
  2004,Li 2005,Wabnig 2005} master equation  of the 
density matrix of the NMR subject to a measurement of a QPC detector and an
influence of a thermal bath. 
Our non-Markovian equation is valid for
arbitrary bath temperatures, electrode reservoir temperatures, bias voltages
and NMR frequencies as long as the approximation used in our approach,
namely the second-order perturbation in the system-bath and
system-reservoir coupling strengths, holds. 
In Sec.~\ref{sec:uncondition_Markovian_ME}, we present the unconditional
non-Markovian master equation and 
show that the unconditional
non-Markovian master equation we obtain reduces, in appropriate limits, to
various Markovian versions of the master equations in the literature. 
In Sec.~\ref{sec:NMR_dynamics}, 
we calculate the non-Markovian expectation values of the NMR dynamical
variables.  The time-dependent transport current through the QPC
which contains information about the measured NMR system is calculated
in Sec.~\ref{sec:current}. We
follow Ref.~\onlinecite{Wabnig 2005} to categorize the non-Markovian average
current into several physically distinct contributions. We find an extra
transient current term that has a substantial contribution to the total
transient current in the Non-Markovian case. Numerical results together with
discussions are presented in Sec.~\ref{sec:numerics}. 
A conclusion is given in Sec.\ref{sec:conclusion}.

\section{Hamiltonian of the NMR-QPC model}
\label{sec:model}

\begin{figure}[tb]
\includegraphics[width=\linewidth]{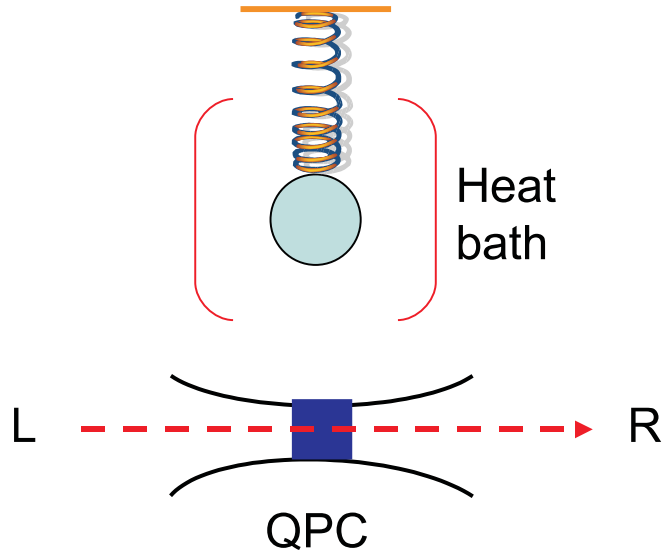}
\caption{(Color online) Schematic diagram of a nanomechanical resonator
(NMR) coupled to a thermal reservoir and measured by a quantum point contact
(QPC) detector.}
\label{fig:NMR-QPC_model}
\end{figure}

In this section, we describe the model of a NMR
that is subject to a measurement by a low-transparency QPC or electric tunnel junction\cite{Mozyrsky 2002,Mozyrsky 2004,Clerk
2004,Wabnig 2005,Smirnov 2003,Wabnig 2007} and is under an influence of a
thermal environment (see Fig.~\ref{fig:NMR-QPC_model}). 
In this model, the NMR is considered as a quantum
harmonic oscillator, and the thermal environment and the QPC electrodes (leads)
are treated as an equilibrium bosonic bath and non-equilibrium fermionic
reservoirs, respectively. By considering the NMR linearly coupled to the QPC,
the Hamiltonian can then be written as 
\begin{equation}
H=H_{S}+H_{B}+H_{I},  \label{totalHamiltonian}
\end{equation}%
where 
\begin{equation}
H_{S}=\frac{p^{2}}{2m}+\frac{1}{2}m\omega _{o}^{2}x^{2},
\label{systemHamiltonian}
\end{equation}
\begin{equation}
H_{B}=H_{\rm leads}+\sum_{n}\left( \frac{p_{n}^{2}}{2m_{n}}+\frac{1}{2}%
m_{n}\omega _{n}^{2}q_{n}^{2}\right)   \label{bathHamiltonian}
\end{equation}%
with
\begin{equation}
H_{\rm leads}=\sum_{l=S,D}H^l_{\rm lead}
=\sum_{l=S,D}\sum_{k}\epsilon _{k}^{l}c_{l,k}^{\dagger }c_{l,k}
\label{bathHamiltonian(a)}
\end{equation}
and%
\begin{equation}
H_{I}=H_{\rm tunneling}+\sum_{n}\lambda _{n}q_{n}x,
\label{interHamiltonian}
\end{equation}%
with
\begin{equation}
H_{\rm tunneling}=\sum_{k,q}(T_{kq}+\chi _{kq}x)c_{S,k}^{\dagger
}c_{D,q}+H.c..
\label{interHamiltonian(a)}
\end{equation}
Here $H_{S}$ represents the Hamiltonian of the NMR system, and $m$ and $\omega
_{o}$ are the mass and the (renormalized) natural frequency of the NMR,
respectively. $H_{B}$ represents the Hamiltonian for the left and right leads
(reservoirs) of the QPC and the thermal (bosonic) bath. 
The thermal bath  in Eq.~(\ref{bathHamiltonian}) consists of a large
number of harmonic oscillators with masses $m_{n}$ and frequencies
$\omega _{n}$, respectively. 
In Eq.~(\ref{bathHamiltonian(a)}), 
$c_{l,k}$ and $\epsilon_{k}^{l}$ are, respectively, the
fermion (electron) reservoir annihilation operators and energies with
wave vector $k$ for the left (source) or right (drain) lead of
the QPC.  The interaction
Hamiltonian $H_{I}$,  Eq.~(\ref{interHamiltonian}),
contains two parts: the first term
describes the tunneling Hamiltonian 
of the electrons through the QPC 
and the second term describes the interaction
between the NMR and the thermal environment. 
In Eq.~(\ref{interHamiltonian(a)}),
the bare tunneling amplitude between respective states with wave
vectors $k$ and $q$ in the
left and right leads (reservoirs) of the QPC is given by $T_{kq}$, and 
$H.c.$ stands for the Hermitian conjugate of the previous term.
So the interaction between the NMR and QPC introduces an effective tunneling
amplitude  \cite{Wabnig 2005} from $T_{kq}\rightarrow T_{kq}+\chi _{kq}x$ in
Eq.~(\ref{interHamiltonian(a)}). 
The NMR and each of the thermal bath oscillators interact bilinearly through
their respective position operators as shown in the last term of 
Eq.~(\ref{interHamiltonian}), and $\lambda _{n}$ is the coupling strength.

\section{Number-resolved quantum master equation}
\label{n-resolvedME}

Non-Markovian dynamics usually means that the current time evolution of the system state depends on its history, and the memory effects typically enter through integrals over the past state history. However, the non-Markovian system dynamics of some class of open quantum system models may be summed up and expressed as a time-local, convolutionless form \cite{Strunz04}
where the dynamics is determined by the system state at the current time $t$ only.  This time-local, convolutionless class of open quantum systems may be treated exactly without any approximation. The quantum Brownian motion model or the damped harmonic oscillator bilinearly coupled to 
a bosonic bath of 
harmonic oscillators \cite{Strunz04,Haake86,Hu92} is a famous example of this class. 
The pure-dephasing spin-boson model
\cite{Palma96,Duan98,Diosi98,Reina02,Schaller08,Goan10}
also belongs to this class. 
The non-Markovian effects in the master equations are taken into
account by the time-dependent decoherence, damping and/or diffusion
coefficients instead of convolution memory integrals.

The perturbative non-Markovian open quantum system theory may also be
categorized into two classes: the time-nonlocal and the time-local (or
time-convolutionless) methods. This has been discussed extensively in
the literature \cite{Breuer2002,Breuer99,Schroder06,Ferraro09}.  
In the quantum
master equation approach, after the Born approximation, the master
equation of the reduced density matrix $\rho(t)$ of the system could be an
integro-differential equation and thus nonlocal in time. In the
interaction picture, the master equation in this case can be written as   
\cite{Breuer2002,Carmichael 1999,Barnett02} 
\begin{eqnarray}
\frac{d\tilde{\rho}(t)}{dt} 
=-\frac{1}{\hbar ^{2}}{\rm Tr}_R\int\nolimits_{0}^{t}dt^{\prime }
\left[ \tilde{H}_{I}\left( t\right) ,\left[ \tilde{H}_{I}
\left( t^{\prime }\right) ,\tilde{\rho}\left( t'\right) \otimes R_0 \right] 
\right],\nonumber \\
\label{tilderho_reduced_pert}  
\end{eqnarray} 
where $\tilde{\rho}(t)$ is the reduced density matrix of the system
and $\tilde{H}_{I}(t)$ is the system-reservoir interaction Hamiltonian in
the interaction picture. In obtaining
Eq.~(\ref{tilderho_reduced_pert}), the assumption of 
the initial total density matrix in the 
uncorrelated (factorized) form of 
$\tilde{\rho}_T(0)=\tilde{\rho}(0)\otimes R_0$ (with $R_0$ being the
reservoir density matrix) and the assumption that the
system-reservoir interaction Hamiltonian satisfies the condition of
\begin{equation}
{\rm Tr}_{R}[\tilde{H}_{I}(t)R_{0}]=0,  
\label{traceless_1st_order}
\end{equation}%
to eliminate the first-order term in $\tilde{H}_{I}$ are made.
However, it can also be shown that another systematically 
perturbative non-Markovian master
equation that is local in time
\cite{Breuer2002,Paz 2001,Breuer99,Schroder06,Ferraro09,Shibata77,Kleinekathofer2004,Liu07,Sinayskiy09,Mogilevtsev09,Haikka10,Ali10}
can be derived from the time-convolutionless
projection operator formalism \cite{Breuer2002,Breuer99,Schroder06,Shibata77} or from the iteration
expansion method \cite{Paz 2001}.  
Under the similar assumptions of the factorized initial system-reservoir
density matrix state and Eq.~(\ref{traceless_1st_order}), the
second-order time-convolutionless master equation can be obtained as  
\begin{eqnarray}
\frac{d\tilde{\rho}(t)}{dt} 
=-\frac{1}{\hbar ^{2}}{\rm Tr}_R\int\nolimits_{0}^{t}dt^{\prime }
\left[ \tilde{H}_{I}\left( t\right) ,\left[ \tilde{H}_{I}
\left( t^{\prime }\right) ,\tilde{\rho}\left( t\right) \otimes R_0 \right] 
\right].\nonumber \\
\label{time_convolutionless_ME}  
\end{eqnarray} 
We note here that obtaining the time-convolutionless 
non-Markovian master equation perturbatively
up to only second order 
in the interaction Hamiltonian 
is equivalent to replacing $\tilde{\rho}(t')$ with $\tilde{\rho}(t)$ 
in Eq.~(\ref{tilderho_reduced_pert}) \cite{Breuer2002,Paz 2001,Breuer99,Schroder06,Ferraro09}. 
One may be tempted to think that the second-order 
time-nonlocal master equation (\ref{tilderho_reduced_pert}) is
more accurate than the second-order time-local (time-convolutionless)
master equation (\ref{time_convolutionless_ME}) since besides the Born approximation, the  (first) Markovian
approximation of replacing $\tilde{\rho}(t')$ with $\tilde{\rho}(t)$
in Eq.~(\ref{tilderho_reduced_pert}) 
seems to be an additional approximation made on the
time-local master equation. But it has been shown
\cite{Breuer2002,Breuer99,Ferraro09,Kleinekathofer2004} that this may not be the
case. In many examples \cite{Breuer2002,Breuer99,Ferraro09,Kleinekathofer2004}, the time-convolutionless approach
works better than the time-nonlocal approach when the exact dynamics
is available and
is used to test the perturbative non-Markovian theory based on these
two approaches. 
The Markovian approximation that we
refer to here 
corresponds to the (second) Markovian approximation where
the bath (reservoir) correlation functions are $\delta$-correlated in time.  
In this (second) Markovian limit, one may change in
Eq.~(\ref{time_convolutionless_ME}) 
the integration variable $t'$ to $\tau=t-t'$    
and then extend 
the upper limit of the time $\tau$ integral  
to infinity (i.e., $t\to \infty$) as 
the bath correlation functions (kernels) are $\delta$-correlated in
time and thus sharply peak at the lower
limit $\tau=0$ of the integral.
Recently, there are several investigations 
\cite{Wolf08,Breuer09,Rivas10,Sun09} of
constructing different 
measures of non-Markovianity to quantify the degree of
non-Markovian behavior of the quantum time evolutions of general systems
in contact with an environment. 
In this article, we do not concern ourselves with
determining the degree of
non-Markovian character as investigated in 
Refs.~\onlinecite{Wolf08,Breuer09,Rivas10,Sun09}. 
The non-Markovian process here refers to  
that we do not make 
the (second) Markovian approximation 
of assuming the bath (reservoir) correlation functions being 
$\delta$-correlated in time
on Eq.~(\ref{time_convolutionless_ME})
to obtain the second-order
time-convolutionless master equation 
so the resultant quantum dynamics of the system state is not Markovian. 
In other words, the influence of the coarse-grained environment
causes nonlocal noise correlations and the memory effects of the
non-Markovian dynamical process are 
contained in the {\em time-dependent}  decoherence, damping and/or diffusion
coefficients of
the time-convolutionless master equation rather than {\em time-independent}
ones in the Markovian case.

We will derive conditional number-resolved (or {\em n}-resolved)
and unconditional quantum master equations for the reduced density matrix of
the NMR up to second order in the effective tunneling amplitude and in
the NMR-thermal-bath coupling strength. The {\em n-}resolved master
equation \cite{Gurvitz 1997,Shnirman98,Krotkov,Goan03,Li 2004,
  Goan04,Clerk 2004,Li 2005,Wabnig 2005} describes the dynamics of the
reduced NMR system state 
conditioned on the number $n$ of the electrons that have tunneled
through the QPC detector in time interval of (0,t), 
and is thus ready to be used to calculate the
transport properties, such as the transport current. The unconditional
master equation can be obtained by summing all possible numbers of electrons 
$n$ in the right lead (drain) of the QPC. 
We will present the derivation of the non-Markovian (time-convolutionless
form) master equation of the reduced density matrix of the NMR system 
by considering only the non-equilibrium QPC fermionic reservoirs 
first, and will include the effect of the equilibrium thermal bosonic
bath into the 
derived master equation later. To proceed the derivation, it is convenient
to go to the interaction picture  \cite{Carmichael 1999,Breuer2002} 
with respect to $H_{0}=H_{S}+H_{\rm leads}$.
The dynamics of the entire system is determined by the time-dependent
tunneling Hamiltonian in
the interaction picture  
\begin{eqnarray}
\tilde{H}_{I}(t)&=&\tilde{H}_{\rm tunneling}(t) \nonumber\\
&=&\sum_{k,q}\left[ T_{kq}+\chi _{kq}x(t)\right] e^{i\left( \epsilon
_{k}^{S}-\epsilon _{q}^{D}\right) t/\hbar }c_{S,k}^{\dagger }c_{D,q}+H.c.,
\label{simply_interaction}
\end{eqnarray}%
where $x(t) =x\cos (\omega _{o}t) +({p}/{m\omega_{o}})\sin(\omega_{o}t)$.
By rewriting 
\begin{equation}
x(t) =\left(\frac{x}{2}
-i\frac{p}{2m\omega _{o}}\right)e^{i\omega _{o}t}+\left(\frac{x}{2}
+i\frac{p}{2m\omega _{o}}\right) e^{-i\omega _{o}t} \, ,
\end{equation}
the interaction (tunneling) Hamiltonian, Eq. (\ref{simply_interaction}), can be
written in the form of 
\begin{equation}
\tilde{H}_{I}(t)=\sum_{k,q}S_{kq}(t)F_{kq}^{^{\dagger
}}(t)+S_{kq}^{\dagger }(t)F_{kq}(t),  \label{responseinteractiveH}
\end{equation}%
where 
\begin{equation}
F_{kq}(t)=e^{-i\left( \epsilon _{k}^{S}-\epsilon _{q}^{D}\right) t/\hbar
}c_{S,k}c_{D,q}^{\dagger }  \label{bath_operator}
\end{equation}%
is the reservoir operator and 
\begin{equation}
S_{kq}(t)=\left[ P_{1}+e^{i\omega _{0}t}P_{2}+e^{-i\omega _{0}t}P_{3}\right]
\label{decom_systemoperator}
\end{equation}%
is an operator in a discrete Fourier decomposition\cite{Stace2004} acting on
the Hilbert space of the NMR system. 
Introducing the dimensionless characteristic length $x_{0}=\sqrt{\hbar
/m\omega _{o}}$ and momentum $p_{0}=\sqrt{m\hbar \omega _{o}}$, 
we may write 
\begin{eqnarray}
P_{1} &=&T_{kq},  \label{P1} \\
P_{2} &=&\tilde{\chi}_{kq}\left( \frac{x}{2x_{0}}-i\frac{p}{2p_{0}}\right) ,
\label{P2} \\
P_{3} &=&\tilde{\chi}_{kq}\left( \frac{x}{2x_{0}}+i\frac{p}{2p_{0}}\right) ,
\label{P3}
\end{eqnarray}%
where $\tilde{\chi}_{kq}=\chi _{kq}x_{0}$ has a dimension the same as
$T_{kq}$. The form of Eq.~(\ref{decom_systemoperator}) indicates that there are
three different jump processes that involve no excitation,
excitation, and relaxation of the energy quanta of the NMR, respectively. 
$P_{1}$ is associated
with elastic (no excitation) tunneling of electrons through the QPC
junction and  
$P_{2}\left( P_{3}\right) $ is associated with inelastic excitation
(relaxation) of electrons tunneling through the QPC with an energy transfer $%
\hbar \omega_{o}$. The energy which relaxes (excites) in response is
provided by the NMR. 
By regarding the tunneling Hamiltonian,
Eq.~(\ref{responseinteractiveH}),
as a perturbative interaction Hamiltonian, 
one can see that the first-order term vanishes, i.e., Eq.~(\ref{traceless_1st_order}) is satisfied, for the density matrix of the QPC reservoirs
(leads) given by $R_0=\rho^S_{\rm lead}\otimes\rho^D_{\rm lead}$, where
\begin{equation}
\rho^l_{\rm lead}=\frac{e^{-\beta(H^l_{\rm lead}-\mu_l \hat{N}_l)}}
{{\rm Tr}_l[e^{-\beta(H^l_{\rm lead}-\mu_l \hat{N}_l)}]}\, ,\quad l=S,D.
\end{equation} 
Here $\hat{N}_l=\sum_{k}c_{l,k}^{\dagger }c_{l,k}$, $\mu_{S}$ and $\mu _{D}$
are the chemical potentials which determine the applied QPC bias
voltage, $eV=\mu _{S}-\mu _{D},$ and $\beta =1/\left( k_{B}T\right)$
is the inverse temperature. 
One may then obtain the second-order (Born
approximation) time-convolutionless non-Markovian 
master equation for the reduced density
matrix of the NMR system by substituting Eq.~(\ref{responseinteractiveH})
into Eq.~(\ref{time_convolutionless_ME}). 
However, in order to make contact of the NMR system with
the QPC detector output current, it will be convenient to obtain an {\em n-}
resolved master equation \cite{Gurvitz 1997,Shnirman98,Krotkov,Goan03,Li 2004,
  Goan04,Clerk 2004,Li 2005,Wabnig 2005}. 
The rotating wave approximation and
Markovian approximation are known to be 
pretty good approximations for systems in quantum optics. 
The rotating wave approximation is a good
approximation provided that the 
strength of the dissipative corrections or the relaxation rate, denoted
generically as $\gamma_R$, is 
small compared to 
the minimum nonzero system frequency difference (energy
difference/$\hbar$) involved in the problem. 
In the present case, this implies that the generic rate $\gamma_R$ 
in conditional $n$-resolved and unconditional master equations should
satisfy the condition of $\gamma_R\ll \omega_0$.
The relevant physical condition for the Markovian approximation is
that the bath correlation time is very small compared to the typical system
response time.
But since the (renormalized) resonant frequency $\omega_0$ of a NMR
is typically in the range of a few hundred KHz to a few GHz which is
much smaller than typical optical frequency  
of $10^{15}$ Hz 
and since the reservoir correlation time in solid-state
devices may not be much shorter than the typical system response time, we
will not make the Markovian approximation 
and the  {\em pre-trace} and {\em post-trace}
rotating wave approximations \cite{Fleming10,Maniscalco04,Intravaia03}
in our derivation of the master equation. 
By first identifying the jump operator terms and 
partially taking trace over the microscopic degrees of freedom of the QPC
reservoirs but keeping track of the number $n$ of electrons that have
tunneled through the QPC detector during the time period $(0,t)$,  and then 
changing from the interaction picture to the Schr\"{o}dinger picture, we can
obtain from  
Eqs.~(\ref{time_convolutionless_ME}) and (\ref{responseinteractiveH})
the time-convolutionless non-Markovian {\em n-}resolved
(conditional) master equation as
\begin{eqnarray}
\dot{\rho}_{R}^{(n)}(t) &=&\frac{1}{i\hbar }[H_{sys},\rho _{R}^{(n)}(t)]-%
\frac{1}{\hbar ^{2}}\int\nolimits_{0}^{t}dt_{1}\text{ }\sum\limits_{k,q;k^{%
\prime },q^{\prime }}\times  \nonumber \\
&&%
\qquad \Biggl\{F_{kq;k^{^{\prime }}q^{\prime }}^{s}(t,t_{1})\Bigl[S_{kq}\left(
t\right) S_{k^{\prime }q^{\prime }}^{\dag }(t_{1})\rho _{R}^{(n)}\left(
t\right) \nonumber \\
&&\qquad \qquad \qquad -S_{k^{^{\prime }}q^{\prime }}^{\dag }(t_{1})\rho
_{R}^{(n+1)}\left( t\right) S_{kq}\left( t\right) \nonumber \\ 
&&\qquad \qquad \qquad -S_{kq}\left(
t\right) \rho _{R}^{(n-1)}\left( t\right) S_{k^{\prime }q^{\prime }}^{\dag
}(t_{1})\nonumber \\ 
&&\qquad \qquad \qquad +\rho _{R}^{(n)}\left( t\right) S_{k^{\prime }q^{\prime }}^{\dag
}(t_{1})S_{kq}\left( t\right) \Bigl]
\nonumber \\
&&\quad \quad \hspace*{0.2cm} +F_{kq;k^{\prime }q^{\prime }}^{a}(t,t_{1})\Bigl[
S_{kq}\left( t\right) S_{k^{\prime }q^{\prime }}^{\dag }(t_{1})\rho
_{R}^{(n)}\left( t\right) \nonumber \\ 
&&\qquad \qquad \qquad -S_{k^{^{\prime }}q^{\prime }}^{\dag }(t_{1})\rho
_{R}^{(n+1)}\left( t\right) S_{kq}\left( t\right)  \nonumber \\
&&\qquad \qquad \qquad +S_{kq}\left(
t\right) \rho _{R}^{(n-1)}\left( t\right) S_{k^{\prime }q^{\prime }}^{\dag
}(t_{1})\nonumber \\ 
&&\qquad \qquad \qquad -\rho _{R}^{(n)}\left( t\right) S_{k^{\prime }q^{\prime }}^{\dag
}(t_{1})S_{kq}\left( t\right) \Bigl]+H.c.\Biggl\}  \nonumber \\
&&+{\cal L}_{\rm damp}[\rho _{R}^{(n)}\left( t\right) ],
\label{n-resolvemastereq}
\end{eqnarray}%
where we have also included the intrinsic dissipation effect of the NMR
induced by interacting with a non-Markovian thermal bosonic environment in the
last term of Eq.~(\ref{n-resolvemastereq}). 
The mode-dependent symmetric and anti-symmetric two-time reservoir
correlation function, $F_{k,q;k^{\prime },q^{\prime }}^{s}(t,t_{1})$ and $%
F_{k,q;k^{\prime },q^{\prime }}^{a}(t,t_{1})$ in 
Eq.~(\ref{n-resolvemastereq}) are respectively 
\begin{eqnarray}
F_{kq;k^{\prime }q^{\prime }}^{s}(t,t_{1})
&\equiv&\frac{1}{2}\left\langle \left\{
F_{k,q}^{^{\dagger }}(t),F_{k^{\prime },q^{\prime }}(t_{1})\right\}
\right\rangle  \nonumber \\
&= &\frac{1}{2}\left\{ N_{Sk}\left( 1-N_{Dq}\right) +\left(
1-N_{Sk}\right) N_{Dq}\right\} \nonumber \\
&& \times e^{i\left( \epsilon _{k}^{S}-\epsilon
_{q}^{D}\right) \left( t-t_{1}\right) /\hbar }\delta _{k,q;k^{\prime
},q^{\prime }},  \label{symmetrykernels}
\end{eqnarray}%
and 
\begin{eqnarray}
F_{k,q;k^{\prime },q^{\prime }}^{a}(t,t_{1})
&\equiv&\frac{1}{2}\left\langle \left[
F_{k,q}^{^{\dagger }}(t),F_{k^{\prime },q^{\prime }}(t_{1})\right]
\right\rangle  \nonumber \\
&=&\frac{1}{2}\left\{ N_{Sk}\left( 1-N_{Dq}\right) -\left(
1-N_{Sk}\right) N_{Dq}\right\} \nonumber \\
&& \times e^{i\left( \epsilon _{k}^{S}-\epsilon
_{q}^{D}\right) \left( t-t_{1}\right) /\hbar }\delta _{k,q;k^{\prime
},q^{\prime }}.  \label{asymmetrykernels}
\end{eqnarray}%
Here the notation $\left\langle \ldots \right\rangle $ indicates the
expectation value over the initial density matrix of the reservoirs
and consequently,  
$F_{k,q;k^{\prime },q^{\prime }}^{s}(t,t_{1})$ and $F_{k,q;k^{\prime
},q^{\prime }}^{a}(t,t_{1})$ are given by the combination of the Fermi
distribution functions $N_{Sk}=\left[ e^{\beta \left( \epsilon_{k}^{S}-\mu
_{S}\right) }+1\right] ^{-1}$ and $N_{Dq}=\left[ e^{\beta \left( \epsilon
_{q}^{D}-\mu _{D}\right) }+1\right] ^{-1}$ of the left (source) and right
(drain) reservoirs of the QPC \cite{Goan01,Goan03,Li 2004,Li 2005}. 

By taking into account relevant tunneling amplitudes and summing over the
wave vectors of the QPC reservoirs, 
the structure of the influence of the QPC
reservoirs on the dynamics of the NMR system is then characterized by the
symmetric and anti-symmetric two-time reservoir correlation kernels, $%
\sum_{k,q,k^{\prime },q^{\prime }}A_{k,q}^{\dag }B_{k,q}F_{k,q;k^{\prime
},q^{\prime }}^{s}\left( t,t_{1}\right) $ and $\sum_{k,q,k^{\prime
},q^{\prime }}A_{k,q}^{\dag }B_{k,q}F_{k,q;k^{\prime },q^{\prime
}}^{a}\left( t,t_{1}\right) ,$ where the value of $A_{k,q}$ and
$B_{k,q}$ could be any one of
the tunneling amplitudes, $T_{k,q}$ and $\tilde{\chi}_{k,q}=\chi _{k,q}x_{0}$.
In the continuous limit, the summation of the QPC reservoir modes can be
replaced by the continuous integrations, $\sum_{k}\sum_{q}%
\rightarrow \int \int d\epsilon _{k}^{S}d\epsilon _{q}^{D}g_{L}(\epsilon
_{k}^S)g_{R}(\epsilon _{q}^D)$, where the energy-dependent densities of states $%
g_{S}(\epsilon _{k}^{S})$ and $g_{D}(\epsilon _{q}^{D})$ are introduced for
left and right QPC electron reservoirs, respectively. In principle, the
tunneling amplitudes, $T_{k,q}=T\left( \epsilon _{k}^{S},\epsilon _{q}^{D}\right) $
and $\tilde{\chi}_{k,q}=\tilde{\chi}\left( \epsilon _{k}^{S},\epsilon
  _{q}^{D}\right) ,$ are also energy-dependent. 
We may deal with any realistic energy function form of the
densities of states and tunneling amplitudes 
 to take into account the memory
effect of the QPC reservoir on the electron transport and the NMR system in
our non-Markovian treatment. For simplicity, we follow several non-Markovian
electron transport studies \cite{Ming2007,Guo05,Guo06,Guo08,Welack
  2006,Jin10,Matisse 2008,Kleinekathofer2004,Wingreen 1994,Ho 2009}
by considering a spectral
density with energy-dependent densities of states and tunneling amplitudes
absorbed in a Lorentzian form as
\begin{eqnarray}
J_{A,B}\left( \epsilon _{k}^{S},\epsilon _{q}^{D}\right) &=&A^{\dag }\left(
\epsilon _{k}^{S},\epsilon _{q}^{D}\right) B\left( \epsilon
_{k}^{S},\epsilon _{q}^{D}\right) g_{L}(\epsilon _{k}^{S})g_{R}(\epsilon
_{q}^{D})  \nonumber \\
&=&\frac{A_{00}^{\dag }B_{00}g_{L}^{0}g_{R}^{0}\Lambda _{e}^{2}}{(\epsilon
_{k}^{S}-\epsilon _{q}^{D}-E_{i})^{2}+\Lambda _{e}^{2}} \, ,
\label{spectraldensity}
\end{eqnarray}%
where the cut-off energy $\Lambda _{e}$ characterizes the width of the
Lorentzian energy-dependent distribution, the parameter $E_{i}$ denotes the
effect of the variation of the QPC junction barrier potential
\cite{Ming2007} due to the
interaction with the NMR, $A_{00},$ $B_{00},$ $g_{L}^{0}$
and $g_{R}^{0}$ are energy-independent tunneling amplitudes and densities of
states near the average chemical potential.  
Physically, this spectral
density of Eq.~(\ref{spectraldensity})
means that given an electron state with a fixed
energy $\epsilon^S_{k}$ in the left lead, the electron can tunnel into the
electron energy states of the right lead with a central energy at $\epsilon
^D_{q}+E_{i}$ and a Lorentzian width $\Lambda _{e}$. In the limit of $%
\Lambda _{e}\rightarrow 0$ and in the absence of the interaction with the
NMR (i.e., $E_{i}=0$), the QPC spectral density, Eq.~(\ref{spectraldensity}),
is proportional to $\delta \left( \epsilon _{k}^{S}-\epsilon _{q}^{D}\right) 
$ that represents the resonant tunneling process. In the opposite case of
the cut-off energy $\Lambda _{e}\rightarrow \infty $, the QPC spectral
density, Eq.~(\ref{spectraldensity}), becomes energy-independent and reduces
to the constant WBL spectral density used in the literature. 
The average
(effective) zero-temperature 
electron tunneling conductances $(G/e^{2})$ through the QPC
barrier in the WBL can be written as 
$(2\pi/\hbar) A_{00}^{\dag}B_{00}g_{L}^{0}g_{R}^{0}$. Compared with the 
energy-dependent spectral density, the WBL one that assumes all electron
states in the left reservoir having equal likelihood to tunnel to all the
electron states in the right reservoir regardless their energies may not be
a very good physical approximation after all.

We note that the dynamical behaviors of the NMR-QPC system are
sensitive to the actual energy dependence and the bandwidth of the QPC
spectral density (which may be different from
the simple Lorentzian form considered here). 
The realistic  
 energy dependence or function form 
of the densities of the states and the tunneling amplitudes 
in the spectral density 
depends on the detailed QPC electronic structure. 
Here we perform a model calculation for the QPC-NMR
system using a simple Lorentzian spectral density to study the influence of
finite bandwidth (cut-off energy) and memory effect on the NMR system dynamics. 
We will show later
in our numerical treatment of the non-Markovian NMR-QPC system that 
for the parameters we choose, when the bandwidth of the Lorentzian
spectral density of 
Eq.~(\ref{spectraldensity}) is about $\Lambda_e\leq 5\hbar \omega_0$,
the time-dependent coefficients, the dynamical variables of the NMR
and the currents through the QPC differ significantly from their Markovian WBL
counterparts. 
This can be understood as follows. 
As discussed earlier, in the limit of $\Lambda _{e}\rightarrow 0$,
only one channel $\epsilon _{k}^S-\epsilon
_{q}^{D}=E_i$ is involved in the 
electron tunneling processes across the QPC barrier.
The opposite limit of a very
large bandwidth, $\Lambda _{e}\gg|\epsilon _{k}^S-\epsilon
_{q}^{D}-E_i|$, then leads to a
channel-mixture
regime \cite{Ming2007} where great portions of all possible
$\epsilon _{k}^{S}\rightleftharpoons\epsilon _{q}^{D}$ transitions of electron
tunneling between the source (left reservoir or lead) and the drain (right
reservoir or lead) are allowed with weight determined by
$J_{A,B}\left( \epsilon _{k}^{S},\epsilon _{q}^{D}\right)$ and 
with randomness coming from electron
scattering determined by the 
band structure associated with the geometry of the metallic
gates in the QPC.
The electron-tunneling processes with more random 
channel mixture will reduce
the QPC reservoir correlation time \cite{Ming2007} and therefore
 suppress the QPC reservoir memory effect on the NMR dynamics.
Thus the non-Markovian processes will become significant
if the QPC electronic structure can be designed or engineered to have 
a spectral density with a narrow bandwidth comparable to the 
(renormalized) resonant frequency of the NMR
as the QPC reservoir correlation time in this case is comparable
to the NMR system response time (see also the discussions 
regarding Figs.~\ref{fig:bath_CF} and \ref{fig:bath_CF2}
in Sec.~\ref{sec:numerics}).
The typical frequency of NMR is in the range of a few hundred KHz to
a few GHz.
Thus the condition of being able to observe a significant
non-Markovian finite-bandwidth behavior of $\Lambda_e\leq 5\hbar
\omega_0$ suggests that the bandwidth of the QPC spectral density should be
in the range of about $1\sim 20$ $\mu$eV.

With the specified spectral density Eq.~(\ref{spectraldensity}) and the help
of Eqs.~(\ref{symmetrykernels}) and (\ref{asymmetrykernels}), one can
rewrite the {\em n-}resolved master equation (\ref{n-resolvemastereq}) into
the following form
\begin{eqnarray}
\dot{\rho}_{R}^{(n)}(t) &=&-\frac{i}{\hbar }[H_{S},\rho _{R}^{(n)}(t)]  \nonumber \\
&&+\frac{g_{L}^{0}g_{R}^{0}}{\hbar}
\biggl\{ \sum_{i=1}^{3} 
f_{F}^{+}( t,eV+\hbar \omega _{i}) \nonumber \\ 
&&\qquad \qquad \times
\left[ P\rho _{R}^{(n-1)}(t)P_{i}^{\dag }-\rho _{R}^{(n)}(t)P_{i}^{\dag }P\right]
\nonumber \\
&&\qquad
-\sum_{i=1}^{3} f_{B}^{+}(t,-eV-\hbar \omega _{i})\nonumber \\ 
&&\qquad \quad \times
\left[ PP_{i}^{\dag }\rho _{R}^{(n)}(t)-P_{i}^{\dag}\rho_{R}^{(n+1)}(t)P \right]
+H.c. \biggl\}\nonumber \\
&&+{\cal L}_{\rm damp}[\rho _{R}^{(n)}(t)].  
\label{n-resolvemaster}
\end{eqnarray}
Here $P_{i}$ is defined in Eqs.~(\ref{P1})--(\ref{P3}), 
$P=\sum_{i=1}^{3}P_{i}=P_{1}+P_{2}+P_{3}$, the values of the frequency $\omega_{i}$
are given by $\omega _{1}=0$, $\omega _{2}=-\omega _{3}=\omega _{o}$,
and $H.c.$ denotes the Hermitian conjugate of all the previous terms
in the curly bracket of Eq.~(\ref{n-resolvemaster}). 
By the change of the new variables, $\omega
_{k}^{S}=\epsilon _{k}^{S}-\mu_S$ and $\omega_{q}^{D}=\epsilon
_{q}^{D}-\mu_D$, the time-dependent coefficients $f_{F(B)}^{\pm}$ in
Eq.~(\ref{n-resolvemaster}) can be written as
following forms: 
\begin{eqnarray}
f_{F}^{+}\left( t,eV\right) &=&\left[ f_{F}^{-}\left( t,eV\right) \right]
^{\dag }  \nonumber \\
&= &\frac{1}{\hbar }\int_{0}^{t}d\tau \int_{-\infty }^{\infty
}\int_{-\infty }^{\infty }d\omega _{k}^{S}d\omega _{q}^{D}\nonumber \\
&& \quad \times \frac{\Lambda
_{e}^{2}}{(\omega _{k}^{S}-\omega _{q}^{D}+eV-E_{i})^{2}+\Lambda _{e}^{2}} 
\nonumber \\
&& \quad \times \frac{1}{e^{\beta \omega _{k}^{S}}+1}\left( 1-\frac{1}{e^{\beta
\omega _{q}^{D}}+1}\right) \nonumber \\
&& \quad \times e^{+i(\omega _{k}^{S}-\omega _{q}^{D}+eV)\tau
/\hbar },  \label{forwardtunneling1}
\end{eqnarray}%
\begin{eqnarray}
f_{B}^{+}\left( t,-eV\right) &=&\left[ f_{B}^{-}\left( t,-eV\right) \right]
^{\dag }  \nonumber \\
&= &\frac{1}{\hbar }\int_{0}^{t}d\tau \int_{-\infty }^{\infty
}\int_{-\infty }^{\infty }d\omega _{k}^{S}d\omega _{q}^{D}\nonumber \\
&& \quad \times \frac{\Lambda
_{e}^{2}}{(\omega _{k}^{S}-\omega _{q}^{D}+eV-E_{i})^{2}+\Lambda _{e}^{2}} 
\nonumber \\
&& \quad \times \left( 1-\frac{1}{e^{\beta \omega _{k}^{S}}+1}\right) \frac{1}{%
e^{\beta \omega _{q}^{D}}+1}\nonumber \\
&& \quad \times e^{+i(\omega _{k}^{S}-\omega _{q}^{D}+eV)\tau
/\hbar },  \label{backwardtunneling1}
\end{eqnarray}%
\begin{eqnarray}
f_{F}^{+}\left( t,eV\pm \hbar \omega _{o}\right) &=&\left[ f_{F}^{-}\left(
t,eV\pm \hbar \omega _{o}\right) \right] ^{\dag }  \nonumber \\
&= &\frac{1}{\hbar }\int_{0}^{t}d\tau \int_{-\infty }^{\infty
}\int_{-\infty }^{\infty }d\omega _{k}^{S}d\omega _{q}^{D}\nonumber \\
&& \quad \times \frac{\Lambda
_{e}^{2}}{(\omega _{k}^{S}-\omega _{q}^{D}+eV-E_{i})^{2}+\Lambda _{e}^{2}} 
\nonumber \\
&& \quad \times \frac{1}{e^{\beta \omega _{k}^{S}}+1}\left( 1-\frac{1}{e^{\beta
\omega _{q}^{D}}+1}\right) \nonumber \\
&& \quad \times e^{+i(\omega _{k}^{S}-\omega _{q}^{D}+eV\pm \hbar
\omega _{o})\tau /\hbar },  \label{forwardtunneling2}
\end{eqnarray}%
and 
\begin{eqnarray}
f_{B}^{+}\left( t,-eV\mp \hbar \omega _{o}\right) &=&\left[ f_{B}^{-}\left(
t,-eV\mp \hbar \omega _{o}\right) \right] ^{\dag }  \nonumber \\
&=&\frac{1}{\hbar }\int_{0}^{t}d\tau \int_{-\infty }^{\infty
}\int_{-\infty }^{\infty }d\omega _{k}^{S}d\omega _{q}^{D}\nonumber \\
&& \quad \times \frac{\Lambda
_{e}^{2}}{(\omega _{k}^{S}-\omega _{q}^{D}+eV-E_{i})^{2}+\Lambda _{e}^{2}} 
\nonumber \\
&& \quad \times \frac{1}{e^{\beta \omega _{q}^{D}}+1}\left( 1-\frac{1}{e^{\beta
\omega _{k}^{S}}+1}\right) \nonumber \\
&& \quad \times e^{+i(\omega _{k}^{S}-\omega _{q}^{D}+eV\pm \hbar
\omega _{o})\tau /\hbar }.  \label{backwardtunnelingrate}
\end{eqnarray}%
Physically, $f_{F}^{\pm }\left( t,eV\right) $ and $f_{B}^{\pm }\left( t,-eV\right) $
describe the memory effects on the NMR system induced by the elastic
electron tunneling processes in the QPC reservoirs with no excitation of the
NMR. The time-dependent coefficients $f_{F}^{\pm }\left( t,eV\pm \hbar \omega
_{o}\right) $ and $f_{B}^{\pm }\left( t,-eV\mp \hbar \omega _{o}\right) $
describe the memory effects on the NMR system caused by the inelastic electron
tunneling processes that involve the NMR excitation and relaxation,
respectively.

The effect of the thermal bosonic environment on the master equation in the
last line of Eq.~(\ref{n-resolvemaster}) 
can be derived also up to second order in system-environment coupling
strength \cite{Breuer2002,Paz 2001,Liu07} and the result is given as 
\begin{eqnarray}
{\cal L}_{\rm damp}[\rho _{R}^{(n)}(t)] &=&-\frac{i}{\hbar }\left[ \frac{M}{2}%
\tilde{\Omega}_{o}^{2}(t) x^{2},\rho _{R}^{(n)}(t)\right] 
\nonumber \\
&&-\frac{i}{\hbar }\gamma _{o}(t) \left[ x,\left\{ p,\rho _{R}^{(n)}(t)\right\} \right]
\nonumber \\
&&-\frac{1}{\hbar ^{2}}D_{o}(t) \left[x,\left[ x,\rho _{R}^{(n)}(t)\right] \right]  \nonumber \\
&&+\frac{1}{\hbar ^{2}}h_{o}(t) \left[ x,\left[ p,\rho _{R}^{(n)}(t)\right] \right] ,
\label{non-Markovian_thermal}
\end{eqnarray}
where the time-dependent coefficients are 
\begin{equation}
\tilde{\Omega}_{o}^{2}(t)=-\frac{2}{m}\int_{0}^{t}d\tau\cos (\omega
_{o}\tau)\eta (\tau),  \label{C10}
\end{equation}%
\begin{equation}
\gamma _{o}(t)=\frac{1}{m\omega _{o}}\int_{0}^{t}d\tau\sin (\omega
_{o}\tau)\eta (\tau),  \label{C11}
\end{equation}%
\begin{equation}
D_{o}(t)=\hbar \int_{0}^{t}d\tau\cos (\omega _{o}\tau)\nu
(\tau),  \label{C12}
\end{equation}%
\begin{equation}
h_{o}(t)=-\frac{\hbar }{m\omega _{o}}\int_{0}^{t}d\tau\sin (\omega
_{o}\tau)\nu (\tau).  \label{C13}
\end{equation}%
Here $\tilde{\Omega}_o^{2}(t)$ is the frequency shift due to the coupling to the
thermal environment, $\gamma _{o}(t)$ is the dissipative coefficient, and $%
D_{o}(t)$ and $h_{o}(t)$ represent the diffusion coefficients. The two
kernels $\eta (\tau)$ and $\nu (\tau)$ appearing in 
Eqs.~(\ref{C10})$-$(\ref{C13}) are so-call dissipation and noise kernels
, respectively, and are defined as 
\begin{equation}
\eta (\tau)=\int_{0}^{\infty }d\omega J\left( \omega \right) \sin
\left( \omega \tau\right) ,  \label{dissipation_corr}
\end{equation}%
and 
\begin{equation}
\nu (\tau)=\int_{0}^{\infty }d\omega J\left( \omega \right) \cos
\left( \omega \tau\right) \coth \left( \beta \hbar \omega /2\right) ,
\label{noise_corr}
\end{equation}%
where 
$J(\omega)$ is the spectral density of the bosonic
environment defined as
\begin{equation}
J\left( \omega \right) =\sum_{n}\frac{\lambda _{n}^{2}}{2m_{n}\omega _{n}}
\delta \left( \omega -\omega _{n}\right).  \label{spectral_func_thermal}
\end{equation}

Again, we could, in principle, deal with any given form of the spectral
density. But as a particular example, we use the following form of spectral
density with a Lorentz-Drude cut-off function to specify the
environment \cite{Breuer2002,Paz 2001,Liu07,Hu92}
\begin{equation}
J\left( \omega \right) =\frac{2}{\pi }m\gamma \omega \left( \frac{\omega }{%
\Lambda }\right) ^{n-1}\frac{\Lambda _{o}^{2}}{\Lambda _{o}^{2}+\omega ^{2}},
\label{simplespectralfunc}
\end{equation}%
where $\Lambda _{o}$ is the cut-off frequency, $\gamma$ is a constant
characterizing the strength of the interaction with the environment, and $m$
is the mass of the NMR. For simplicity, we will take the commonly used spectral
density of an Ohmic bath, i.e., $n=1$ case in Eq.~(\ref{simplespectralfunc}).

The {\em n}-resolved master equation (\ref{n-resolvemaster}) with Eq.~(\ref%
{non-Markovian_thermal}) was derived without making the Markovian and the
pre-trace and post-trace rotating wave approximations, and the only
approximations we use are the
second-order perturbation theory, 
the initially factorized system-bath density matrix, and the
forms of the spectral densities of Eqs.~(\ref{spectraldensity}) and
(\ref{simplespectralfunc}). So the {\em n}-resolved master equation 
is valid for arbitrary bias voltages and environment of temperature, as long
as the perturbation theory that we use up to second order in the
system-QPC and system-environment coupling strength holds.

\section{Unconditional master equation and Markovian limit}
\label{sec:uncondition_Markovian_ME}

\subsection{Unconditional master equation}
\label{sec:unconditional_ME}

In this subsection, we present the unconditional master equation for the
reduced density matrix of the NMR system. Statistically, the unconditional
master equation can be straightforwardly obtained by summing up Eq.~(\ref%
{n-resolvemaster}) over all possible electron number {\em n, }i.e., $\rho
_{R}(t)=\sum\nolimits_{n}\rho _{R}^{\left( n\right) }.$ Despite the
different nature between the non-equilibrium fermionic QPC reservoir and the
thermal bosonic environment, by combining the relevant terms together, the
unconditional non-Markovian master equation can be cast into a simple form
similar to the non-Markovian quantum Brownian motion master equation as 
\begin{eqnarray}
\dot{\rho}_{R}(t) &=&-\frac{i}{\hbar }\left[ H_{sys}+\frac{m}{2}%
\left( \tilde{\omega}_{e}^{2}(t)+\tilde{\Omega}_{o}^{2}(t)\right) x^{2},\rho
_{R}(t)\right]  \nonumber \\
&&-\frac{i}{\hbar }\left( \gamma _{e}(t)+\gamma _{o}(t)\right) %
\left[ x,\left\{ p,\rho _{R}(t)
\right\} \right]  \nonumber \\
&&-\frac{1}{\hbar ^{2}}\left( D_{e}(t)+D_{o}(t)\right) \left[ 
x,\left[ x,\rho _{R}(t)\right] \right]  \nonumber \\
&&+\frac{1}{\hbar ^{2}}\left( h_{e}(t)+h_{o}(t)\right) \left[ 
x,\left[ p,\rho _{R}(t)\right] \right] .
\label{quantummasterequation}
\end{eqnarray}%
The whole non-Markovian character of the dynamics of the NMR system is
contained in the time-dependent coefficients appearing in the master
equation. The time-dependent coefficients that come from the QPC electron
reservoirs are denoted with a subscript $e$. The frequency renormalization $%
\tilde{\omega}_{e}^{2}(t)$, the damping coefficient $\gamma _{e}(t)$, the
decoherence coefficient $D_{e}(t)$ and the diffusion coefficient $h_{e}(t)$
are, respectively, given by 
\begin{equation}
\tilde{\omega}_{e}^{2}(t)=\frac{\hbar G_{xx}}{\pi m}\mathrm{Im}\left[ \xi
_{1}^{a}\left( t\right) +\xi _{2}^{a}\left( t\right) \right] ,
\label{renormalization}
\end{equation}%
\begin{equation}
\gamma _{e}(t)=\frac{\hbar G_{xx}}{2\pi m\omega _{o}}\mathrm{Re}\left[ \xi
_{1}^{a}\left( t\right) -\xi _{2}^{a}\left( t\right) \right] ,
\label{damping}
\end{equation}%
\begin{equation}
D_{e}(t)=\frac{\hbar ^{2}G_{xx}}{2\pi }\mathrm{Re}\left[ \xi _{1}^{s}\left(
t\right) +\xi _{2}^{s}\left( t\right) \right] ,  \label{decoherent}
\end{equation}%
\begin{equation}
h_{e}(t)=\frac{\hbar ^{2}G_{xx}}{2\pi m\omega _{o}}\mathrm{Im}\left[ \xi
_{1}^{s}\left( t\right) -\xi _{2}^{s}\left( t\right) \right],
\label{diffusion}
\end{equation}%
where 
\begin{eqnarray}
G_{xx} &=&\frac{2\pi }{\hbar }g_{L}^{0}g_{R}^{0}\left\vert \chi
_{00}\right\vert ^{2}, \\
\xi _{1}^{s}\left( t\right) &=&f_{F}^{+}\left( t,eV+\hbar \omega _{0}\right)
+f_{B}^{+}\left( t,-eV-\hbar \omega _{o}\right) ,  \label{xi1s} \\
\xi _{1}^{a}\left( t\right) &=&f_{F}^{+}\left( t,eV+\hbar \omega _{o}\right)
-f_{B}^{+}\left( t,-eV-\hbar \omega _{o}\right) ,  \label{xi1a} \\
\xi _{2}^{s}\left( t\right) &=&f_{F}^{+}\left( t,eV-\hbar \omega _{o}\right)
+f_{B}^{+}\left( t,-eV+\hbar \omega _{o}\right) ,  \label{xi2s} \\
\xi _{2}^{a}\left( t\right) &=&f_{F}^{+}\left( t,eV-\hbar \omega _{o}\right)
-f_{B}^{+}\left( t,-eV+\hbar \omega _{o}\right).  \label{xi2a}
\end{eqnarray}

We note that due to the interaction with the thermal environment,
the frequency shift term $\tilde{\Omega}_{o}^{2}(t)$ in Eq. (\ref%
{quantummasterequation}) diverges as the cut-off frequency $\Lambda_o
\rightarrow \infty $ and thus is not physical\cite{Breuer2002,Paz
2001,Liu07,Hu92}. Therefore, a regularization procedure\cite%
{Ryder96,Mandl93} is needed for the frequency renormalization. We adopt the
view of the renormalization \cite{Ryder96,Mandl93} to regard the
frequency in the original Hamiltonian as a finite renormalized frequency $%
\omega _{o}$ and add a frequency counter term with a frequency \cite%
{Breuer2002,Paz 2001,Liu07,Hu92} defined as 
\begin{equation}
\Omega _{c}^{2}=\frac{1}{m}\sum_{n}\frac{\lambda _{n}^{2}}{m_{n}\omega
_{n}^{2}}=\frac{2}{m}\int_{0}^{\infty }d\omega \frac{J(\omega )}{\omega }
\end{equation}
to cancel at large times the frequency shift $\tilde{\Omega}_{o}^{2}(t).$
Similar to this reasoning, another frequency counter term with a
frequency $\omega_c^{2}$ is introduced \cite{Wabnig 2005} to 
compensate at long
times the frequency shift $\tilde{\omega}_{e}^{2}(t)$ induced as a result of
the coupling to the QPC leads (reservoirs). The physical frequency in this case
is then $\omega _{p}^{2}(t)=\omega _{o}^{2}+\tilde{\omega}_{e}^{2}(t)+\tilde{%
\Omega}_{o}^{2}(t)+\omega_c^{2}+\Omega _{c}^{2}$ and approaches the
finite renormalized frequency $\omega _{o}$ at large times. The
non-Markovian master equations (\ref{n-resolvemaster}) and (\ref%
{quantummasterequation}) are the main results of this paper.

\subsection{Markovian limit}
\label{sec:Markovian}

Next, we show that by taking appropriate limits, our
non-Markovian master equations can recover the various Markovian
master equations 
reported in the literature. The Markovian approximation is valid
when the bath correlation time is much smaller than the characteristic time
scale of the system of interest. The bath correlation time is determined by
the bath correlation kernels (functions) and depends on the form of the
bath spectral density. We will perform the numerical calculation of the QPC
reservoir correlation time in the next section to investigate how the
reservoir correlation time is varied as a function of various parameters in
the problem.

Here, if we nevertheless take the Markovian approximation of very short bath
correlation times of the QPC reservoirs and of the thermal bosonic
environment, 
this is equivalent to assuming that the bath correlation functions
(kernels) are $\delta-$correlated in time and thus the upper limit $t$
of the time $\tau$ integrals in 
Eqs.~(\ref{forwardtunneling1})-(\ref{backwardtunnelingrate}) for QPC
reservoirs and in 
Eqs.~(\ref{C10})-(\ref{C13}) for the thermal bosonic bath can be taken to $%
t\rightarrow \infty $. Another commonly used assumption in the Markovian
limit is the so-called WBL approximation. This assumption may not seem
essential to evaluate the integrations if one already makes the very short
bath correlation time approximation. But the assumption of very short bath
correlation times can be justified in various models of the bath spectral
densities with very large cut-off energies. This has been demonstrated for
the bosonic Ohmic bath \cite{Breuer2002,Paz 2001,Liu07,Hu92}. We will
show in Sec.~\ref{sec:numerics} that this is also the case for the simple
Lorentzian spectral density of Eq.~(\ref{spectraldensity}) for the
non-equilibrium QPC fermionic reservoirs. Thus, if we take the Markovian
approximation of very short correlation times (integration limit $%
t\rightarrow \infty $), 
then the time-dependent coefficients in the master equations (\ref%
{n-resolvemaster}), (\ref{non-Markovian_thermal}) and (\ref%
{quantummasterequation}) become time-independent. Specifically, using the
relation 
\begin{equation}
\lim_{t\rightarrow \infty }\int_{0}^{t}d\tau e^{i\left( \omega -\omega
_{o}\right) \tau }=\pi \delta \left( \omega -\omega _{0}\right) +iPV\left( 
\frac{1}{\omega -\omega _{o}}\right) ,  \label{Integration_formation}
\end{equation}%
where $PV$ indicates the Cauchy principle value, the coefficients 
$f_{F(B)}^{\pm }$ in 
Eqs.~(\ref{forwardtunneling1})$-$(\ref{backwardtunnelingrate}) coming
from the QPC reservoirs can be written as 
\begin{equation}
\lim_{t\rightarrow \infty }f_{F(B)}^{\pm }\left( t,y\right) \equiv
W_{F\left( B\right) }\left[ y\right] \pm i\Theta _{F(B)}\left[ y\right] ,
\label{Markovian_limit}
\end{equation}%
where 
\begin{eqnarray}
W_{F}\left[ y\right]  &=&\pi \frac{y}{1-e^{-y/k_{B}T}}\frac{\Lambda _{e}^{2}%
}{(eV-y+E_{i})^{2}+\Lambda _{e}^{2}},  \label{Markovian_real_F} 
\end{eqnarray}
\begin{eqnarray}
W_{B}\left[ y\right]  &=&\pi \frac{y}{1-e^{-y/k_{B}T}}\frac{\Lambda _{e}^{2}%
}{(eV+y+E_{i})^{2}+\Lambda _{e}^{2}},  \label{Markovian_real_B} 
\end{eqnarray}
\begin{eqnarray}
\Theta _{F}\left[ y\right]  &=&\int_{-\infty }^{\infty }\int_{-\infty
}^{\infty }d\omega _{k}^{S}d\omega _{q}^{D}
\frac{\Lambda _{e}^{2}
}{(\omega_k^S-\omega_k^D-E_{i})^{2}+\Lambda _{e}^{2}}
\nonumber \\
&& \quad \times \frac{1}{e^{\beta \omega _{k}^{S}}+1}\text{ }\left( 1-\frac{1}{%
e^{\beta \omega _{q}^{D}}+1}\right) \nonumber \\
&& \quad \times PV\left( \frac{1}{\omega _{k}^{S}-\omega
_{q}^{D}+eV+y}\right) ,  \label{Markovian_image_F} 
\end{eqnarray}
\begin{eqnarray}
\Theta _{B}\left[ y\right]  &=&\int_{-\infty }^{\infty }\int_{-\infty
}^{\infty }d\omega _{k}^{S}d\omega _{q}^{D}
\frac{\Lambda _{e}^{2}
}{(\omega_k^S-\omega_k^D-E_{i})^{2}+\Lambda _{e}^{2}}
\nonumber \\
&&\quad \times \frac{1}{e^{\beta \omega _{q}^{D}}+1}\text{ }\left( 1-\frac{1}{%
e^{\beta \omega _{k}^{S}}+1}\right) \nonumber \\
&& \quad \times PV\left( \frac{1}{\omega _{k}^{S}-\omega
_{q}^{D}+eV-y}\right).  \label{Markovian_image_B}
\end{eqnarray}%
We note here that we separate the Markovian approximation from the WBL
approximation although the Markovian approximation can often be justified by
considering a very large cut-off energy. So the cut-off energy $\Lambda _{e}$
remains in Eqs.~(\ref{Markovian_real_F})--(\ref{Markovian_image_B}). 
If the WBL ($\Lambda
_{e}\rightarrow \infty $) is taken, then the real parts of $f_{F(B)}^{\pm
}\left( t\rightarrow \infty ,y\right) $ when multiplied by the factor $%
2g_{L}^{0}g_{R}^{0}A_{00}B_{00}/\hbar $ become
the finite-temperature forward (backward) Markovian WBL electron tunneling
rates of $\Gamma _{AB}^{MT}(y)=(2\pi/\hbar)
g_{L}^{0}g_{R}^{0}A_{00}B_{00}/[1-\exp (-y/k_{B}T)]$
in the literature \cite{Goan01,Goan03,Stace2004,Li 2004,Goan04,Li
2005,Mozyrsky 2002,Wabnig 2005,Clerk 2004}, where the value of
$A_{00}$ and $B_{00}$ could be either one of the tunneling amplitudes
$T_{00}$ and $\tilde{\chi}_{00}$. 
As a result, in the Markovian limit, the
frequency renormalization, the damping coefficient, the decoherence
coefficient, and the diffusion coefficient in
Eqs.~(\ref{renormalization})-(\ref{diffusion}) due to the QPC
reservoirs in the WBL ($\Lambda _{e}\rightarrow \infty $) become, respectively
\begin{eqnarray}
\left( \tilde{\omega}_{e}^{M}\right) ^{2} &=&\frac{\hbar G_{xx}}{\pi m}%
\left\{ \Theta _{F}\left[ eV+\hbar \omega _{o}\right] -\Theta _{B}\left[
-eV-\hbar \omega _{o}\right] 
\right. \nonumber \\
&&\quad \quad  \left. 
+\Theta _{F}\left[ eV-\hbar \omega _{o}\right]
-\Theta _{B}\left[ -eV+\hbar \omega _{o}\right] \right\} ,  \label{Mar_renor}
\\
\gamma _{e}^{M} &=&\frac{\hbar ^{2}}{m}G_{xx},  \label{Mar_damping} \\
D_{e}^{M} &=&\frac{\hbar ^{2}G_{xx}}{2}\left[ \left( eV+\hbar \omega
_{o}\right) \coth \frac{eV+\hbar \omega _{o}}{2k_{B}T} \right. \nonumber \\
&&\qquad \qquad \left.
+\left( eV-\hbar
\omega _{o}\right) \coth \frac{eV-\hbar \omega _{o}}{2k_{B}T}\right], 
\label{Markov_decoherence} \\
h_{e}^{M} &=&\frac{\hbar G_{xx}}{2\pi m\omega _{o}}\left\{ \Theta _{F}\left[
eV+\hbar \omega _{o}\right] +\Theta _{B}\left[ -eV-\hbar \omega _{o}\right]
\right. \nonumber \\
&&\qquad \left.
-\Theta _{F}\left[ eV-\hbar \omega _{o}\right] -\Theta _{B}\left[ -eV+\hbar
\omega _{o}\right] \right\}.
\end{eqnarray}%
Similarly, the frequency renormalization, the damping coefficient and the
diffusion coefficients in Eqs.~(\ref{C10})-(\ref{C13}) due to the Ohmic
thermal environment in the Markovian WBL ($\Lambda _{o}\rightarrow \infty $)
can also be obtained as \cite{Breuer2002} 
\begin{eqnarray}
\left( \tilde{\Omega}_o^{M}\right) ^{2} 
&=&-\frac{1}{m}\sum\limits_{n}\frac{\lambda_{n}^{2}}{m_{n}\omega _{n}^{2}}
=-\frac{2}{m}\int\nolimits_{0}^{\infty }d\omega \frac{%
J\left( \omega \right) }{\omega }=-\Omega _{c}^{2}, \nonumber \\
\label{Mar_renor_bosonic} \\
\gamma _{o}^{M} &=&\gamma ,  \label{Mar_damping_bosonic} \\
D_{o}^{M} &=&m\gamma \hbar \omega _{o}\coth \left( \frac{\hbar \omega _{o}}{%
2k_{B}T}\right) ,  \label{Mar_decoherent_bosonic} \\
h_{o}^{M} &=&\gamma k_{B}T\sum\limits_{n=-\infty }^{\infty }\frac{-\nu _{n}}{%
\left( \nu _{n}^{2}+\omega _{o}^{2}\right) },  \label{Mar_diffusion_bosonic}
\end{eqnarray}%
where $\upsilon _{n}={2\pi nk_{B}T}/{\hbar }$ are known as the Matsubara
frequencies.

We note that our unconditional non-Markovian master equation (\ref%
{quantummasterequation}) in the Markovian WBL recovers the Markovian master
equation (2.12) in Ref.~\onlinecite{Wabnig 2005}. In the special case of the
zero-temperature and high-bias limit where $D_{e}^{M}=m\gamma _{e}^{M}eV$,
we recover Eq.~(6) of Ref.~\onlinecite{Mozyrsky 2002} if the coefficients coming
from the contributions of the thermal bosonic bath are neglected. Similarly,
the conditional non-Markovian {\em n-}resolved master equation (\ref%
{n-resolvemaster}) also reduces to the Markovian {\em n-}resolved master
equation (2.9) in Ref.~\onlinecite{Wabnig 2005}. Taking the
high-temperature limit of the bosonic environment and
Fourier-transforming in the {\em n} index, the conditional 
{\em n-}resolved master equation (\ref{n-resolvemaster}) in the Markovian
limit also reduces to Eq.~(2) of Ref.~\onlinecite{Clerk 2004} 
if the dc bias case is considered and the transmission phase $\eta $ is set to
zero in Ref.~\onlinecite{Clerk 2004}. Again, considering only the QPC
reservoirs and in the special case of the zero-temperature and high-bias
limits, one obtains Eq.~(5) of Ref.~\onlinecite{Mozyrsky 2002} in the Markovian
limit from the conditional {\em n-}resolved master equation
(\ref{n-resolvemaster}). 

\section{Dynamics of the NMR}
\label{sec:NMR_dynamics}

Using the master equation (\ref{quantummasterequation}), we can obtain the
equation of motion for the mean or expectation value of any physical
operation $O$ of the NMR by calculating $\frac{d\langle O\rangle }{dt}=%
\mathrm{Tr}[O\dot{\rho}_{R}(t)]$. So the equations of motion of the mean
(expectation value) of the position and the momentum are%
\begin{eqnarray}
\frac{d\left\langle x\left( t\right) \right\rangle }{dt} &=&\frac{%
\left\langle p\left( t\right) \right\rangle }{m},  \label{x_rep} \\
\frac{d\left\langle p\left( t\right) \right\rangle }{dt} &=&-m\text{ }\omega
_{p}^{2}(t)\left\langle x\left( t\right) \right\rangle \nonumber \\
&& -2\left[ \gamma
_{e}(t)+\gamma _{o}\left( t\right) \right] \left\langle p\left( t\right)
\right\rangle ,  \label{p_rep}
\end{eqnarray}%
and for the second moments we obtain 
\begin{eqnarray}
\frac{d\left\langle x^{2}\left( t\right) \right\rangle }{dt} &=&\frac{1}{m}%
\left\langle \left\{ x,\text{ }p\right\} \left( t\right) \right\rangle ,
\label{square_x} \\
\frac{d\left\langle p^{2}\left( t\right) \right\rangle }{dt} &=&-m\text{ }%
\omega _{p}^{2}(t)\left\langle \left\{ x,\text{ }p\right\} \left( t\right)
\right\rangle \nonumber \\
&& -4\left[ \gamma _{e}(t)+\gamma _{o}(t)\right] \left\langle
p^{2}\left( t\right) \right\rangle  \nonumber \\
&&+2\left[ D_{e}(t)+D_{o}(t)\right] ,  \label{square_p} \\
\text{ }\frac{d\left\langle \left\{ x,p\right\} \left( t\right)
\right\rangle }{dt} &=&2\frac{\left\langle p^{2}\left( t\right)
\right\rangle }{m}-2m\omega _{p}^{2}(t)\left\langle x^{2}\left( t\right)
\right\rangle \nonumber \\
&& -2\left[ \gamma _{e}(t)+\gamma _{o}(t)\right] \left\langle
\left\{ x,\text{ }p\right\} \left( t\right) \right\rangle  \nonumber \\
&&+2\left[ h_{e}(t)+h_{o}(t)\right] .  \label{square_xp}
\end{eqnarray}
Combining Eqs.~(\ref{x_rep}) and (\ref{p_rep}) yields 
\begin{equation}
\frac{d^{2}\langle x\left( t\right) \rangle }{dt^{2}}+2\gamma _{tot}\left(
t\right) \frac{d\langle x\left( t\right) \rangle }{dt}+\omega
_{p}^{2}(t)\langle x\left( t\right) \rangle =0,  \label{classical_position}
\end{equation}%
where $\gamma_{tot}=\gamma_e+\gamma_o$.

One may in principle solve the time evolutions of the
differential equations (\ref{x_rep})--(\ref{square_xp}) 
and the numerical results will be presented
in Sec.~\ref{sec:numerics}. 
Simple analytical expressions of the steady-state ($t\rightarrow \infty $) 
solutions can, however, be obtained as \cite{Sandulescu1987} 
\begin{eqnarray}
\left\langle x\right\rangle _{t\rightarrow \infty } &=&\left\langle
p\right\rangle _{t\rightarrow \infty }=0, 
\label{steady_state0}\\
\left\langle \left\{ x,p\right\} \right\rangle _{t\rightarrow \infty } &=&0,
\label{steady_state1} \\
\left\langle x^{2}\right\rangle _{t\rightarrow \infty }
&=&\lim_{t\rightarrow \infty }\frac{1}{2m\omega _{p}^{2}(t)}\biggl( \frac{%
D_{e}(t)+D_{o}(t)}{m\left[ \gamma _{e}(t)+\gamma _{o}(t)\right] } 
\nonumber \\
&& \qquad \qquad +2\left[ h_{e}(t)+h_{o}(t)\right] \biggl) ,  \label{steady_state2} \\
\left\langle p^{2}\right\rangle _{t\rightarrow \infty }
&=&\lim_{t\rightarrow \infty }\frac{D_{e}(t)+D_{o}(t)}{2\left[ \gamma
_{e}(t)+\gamma _{o}(t)\right] }\, .  \label{steady_state3}
\end{eqnarray}
For the moment, let us 
consider the case where the influence of the thermal
environment is neglected. 
We also note that, for typical values of 
finite electric reservoir temperatures and finite
electric bias voltages, the diffusion coefficient $h_{e}(t)/\hbar$ is generally
much smaller than $D_{e}(t)/p_0^2$ and $\gamma_{e}(t)$ and thus is
often neglected. In this case, 
we obtain from Eqs.~(\ref{steady_state2}), (\ref{Mar_damping})
and (\ref{Markov_decoherence}) the steady-state 
$\langle x^{2}\rangle _{t\to\infty}$ in the WBL as 
\begin{eqnarray}
\left\langle x^{2}\right\rangle _{t\to\infty}
&=&\frac{1}{4m\omega _{o}^{2}}
\left[ \left( eV+\hbar \omega
_{o}\right) \coth \frac{eV+\hbar \omega _{o}}{2k_{B}T}
\right. \nonumber \\
&& \left.  +\left( eV-\hbar
\omega _{o}\right) \coth \frac{eV-\hbar \omega _{o}}{2k_{B}T}\right] 
.  \label{stationaryxx}
\end{eqnarray}%
At zero temperature ($k_{B}T=0$) and low voltages ($eV\ll \hbar \omega _{o}$%
), we have $\left\langle x^{2}\right\rangle \approx \frac{\hbar }{2m\omega
_{o}}$. In this case, the NMR is in the ground state and is
independent of the bias voltage as the bias voltage is unable to excite the
NMR from its ground state. On the other hand, at high voltages ($%
eV\gg \hbar \omega _{o}$), the NMR is no longer in the ground state
and $\left\langle x^{2}\right\rangle \approx \frac{eV}{2m\omega _{o}^{2}}$%
. \cite{Wabnig 2005} At high temperatures ($k_{B}T\gg eV,$ $\hbar \omega _{o}$%
), the quantum mean square of the position of the NMR becomes $%
\left\langle x^{2}\right\rangle \approx \frac{k_{B}T}{m\omega _{o}^{2}}$
that is expected from a classical oscillator in thermal equilibrium.

\section{Transport current }
\label{sec:current}



With the {\em n}-resolved time-convolutionless master equation for $\rho
_{R}^{(n)}(t)$, one is readily able to compute 
transport current $I\left( t\right) =e\frac{d\left\langle N\left( t\right)
\right\rangle }{dt},$ where $\left\langle N\left( t\right) \right\rangle
=\sum_{n}nP\left( n,t\right) =\sum_{n}n\mathrm{Tr}\left[ \rho _{R}^{(n)}(t)%
\right] $ is the expectation value of the number of electrons that have
tunneled into the right lead (drain) in time $t$. Here $\mathrm{Tr}$ means
tracing the density matrix over the degrees of freedom of the NMR system. 
Inserting Eq.~(\ref{n-resolvemaster}) into $I\left( t\right)
=e\sum\nolimits_{n}n\mathrm{Tr}\left[ \dot{\rho}_{R}^{(n)}(t)\right] $ 
gives rise to \cite{Gurvitz 1997,Shnirman98,Krotkov,Goan03,Li 2004,Goan04,Clerk
  2004,Li 2005,Wabnig 2005} 
\begin{eqnarray}
\frac{I\left( t\right) }{e}&=&
\frac{g_{L}^{0}g_{R}^{0}}{\hbar }\sum_{i=1}^{3}%
\mathrm{Tr}\left\{ \left[ f_{F}^{+}\left( t,eV+\hbar \omega _{i}\right)
P_{i}^{\dag }P\rho _{R}(t) \right.\right. \nonumber \\
&& \left. \left.  -f_{B}^{+}\left( t,-eV-\hbar \omega _{i}\right)
PP_{i}^{\dag }\rho _{R}(t)\right] +H.c.\right\} .  \label{current_operator}
\end{eqnarray}%
Here $\rho _{R}(t)=\sum\nolimits_{n}\rho _{R}^{(n)}(t)$ is the unconditional
density matrix of the NMR. This non-Markovian average current is valid
for arbitrary 
QPC lead temperatures and arbitrary bias voltages as long as the
second-order perturbation theory holds. 
We follow Ref.~\onlinecite{Wabnig 2005} to categorize the non-Markovian average
current, Eq.~(\ref{current_operator}), into four physically distinct
contributions. Using the definition of $P=\sum_{i=1}^3 P_i$ and the
definition of $P_i$ in 
Eqs.~(\ref{P1})--(\ref{P3}), we write the non-Markovian average
current, Eq.~(\ref{current_operator}), as
\begin{equation}
\frac{I(t)}{e}=\frac{I_{\rm position}(t)}{e} +\frac{I_{P}( t)}{e} +\frac{I_{QM}( t)}{e}
+\frac{I_{\left\{ X,P\right\} }( t)}{e} .  \label{current}
\end{equation}%
The first term in Eq.~(\ref{current}) depends on the oscillation
position of the NMR and can be written as 
\begin{eqnarray}
\frac{I_{\rm position}( t)}{e} &=&\frac{1}{2\pi }\Delta _{1}\left( t\right) \left[
G_{0}+G_{x}\left\langle x\left( t\right) \right\rangle \right] 
\nonumber \\
&&+\frac{1}{4\pi }
\left[\Delta_{2}\left( t\right)+\Delta _{3}\left( t\right)\right]\nonumber \\
&& \quad \times
 \left[ G_{x}\left\langle x\left( t\right) \right\rangle
+G_{xx}\left\langle x^{2}\left( t\right) \right\rangle \right].
\label{x-current}
\end{eqnarray}%
It reduces in the Markovian WBL to the so-called 
Ohmic-like part of the current 
proportional to the conductance defined in 
Ref.~\onlinecite{Wabnig 2005}. 
In Eq.~(\ref{x-current}), the term with conductance 
$G_{0}=\frac{2\pi }{\hbar }g_{L}^{0}g_{R}^{0}|T_{00}|^{2}$
represents the current through the isolated QPC junction, and the remaining
terms due to the coupling to the NMR with 
conductances $G_{x}=\frac{2\pi }{\hbar }g_{L}^{0}g_{R}^{0}\mathrm{Re}[
T_{00}^{\dag }\chi _{00}] $ and $G_{xx}=\frac{2\pi }{\hbar }%
g_{L}^{0}g_{R}^{0}|\chi _{00}|^{2}$  contribute to the 
nonlinear part of
current-voltage characteristic \cite{Rammer2004,Wabnig 2005}
as the state of the NMR will depend on the bias voltage. 
The time-dependent coefficients $\Delta_i(t)$ in the non-Markovian region 
can be written as 
\begin{eqnarray}
\Delta _{1}\left( t\right)  &=&f_{F}^{+}\left( t,eV\right) -f_{B}^{+}\left(
t,-eV\right) \nonumber \\
&& +f_{F}^{-}\left( t,eV\right) -f_{B}^{-}\left( t,-eV\right)  
\nonumber \\
&=&2\mathrm{Re}\left[ f_{F}^{+}\left( t,eV\right) -f_{B}^{+}\left(
t,-eV\right) \right] ,  \label{coeff_electric1} \\
\Delta _{2}\left( t\right)  &=&f_{F}^{+}\left( t,eV+\hbar \omega _{o}\right)
-f_{B}^{+}\left( t,-eV-\hbar \omega _{o}\right) \nonumber \\
&& +f_{F}^{-}\left( t,eV+\hbar
\omega _{o}\right) -f_{B}^{-}\left( t,-eV-\hbar \omega _{o}\right)   \nonumber
\\
&=&2\mathrm{Re}\left[ \xi _{1}^{a}(t)\right] ,  \label{coeff_electric2} \\
\Delta _{3}\left( t\right)  &=&f_{F}^{+}\left( t,eV-\hbar \omega _{o}\right)
-f_{B}^{+}\left( t,-eV+\hbar \omega _{o}\right) \nonumber \\
&& +f_{F}^{-}\left( t,eV-\hbar
\omega _{o}\right) -f_{B}^{-}\left( t,-eV+\hbar \omega _{o}\right)   \nonumber
\\
&=&2\mathrm{Re}\left[ \xi _{2}^{a}(t)\right] ,  \label{coeff_electric3}
\end{eqnarray}%
where $\xi _{1}^{a}(t)$ and $\xi _{2}^{a}(t)$ are defined in Eqs.~(\ref{xi1a}) and
(\ref{xi2a}), respectively. 
The second term in Eq.~(\ref{current}), $I_{P}(t)/e $, is
proportional to the average velocity of the oscillator 
\begin{equation}
\frac{I_{P}( t)}{e} =\frac{G_{p}}{4\omega _{o}\pi m}\left[ \Delta
_{2}\left( t\right) -\Delta _{3}\left( t\right) \right] \left\langle p\left(
t\right) \right\rangle .  \label{p-current}
\end{equation}%
This term is nonvanishing only for an asymmetric junction \cite{Wabnig 2005}%
, $G_{p}=\frac{2\pi }{\hbar }g_{L}^{0}g_{R}^{0}\mathrm{Im}[
T_{00}^{\dag }\chi _{00}] \neq 0$. 
The third term in Eq.~(\ref{current}), $I_{QM}\left( t\right) $,
derived from the commutator of position and momentum operators,
is referred to as the quantum correction to the current \cite{Wabnig 2005}  
\begin{equation}
\frac{I_{QM}( t)}{e} =-
\frac{\hbar G_{xx}}{8\pi m\omega _{o}}\left[ \Delta
_{4}\left( t\right) -\Delta _{5}\left( t\right) \right] .  \label{QM-current}
\end{equation}%
Here the time-dependent coefficients $\Delta _{4}(t)$ and $\Delta_{5}(t)$ 
can be written as 
\begin{eqnarray}
\Delta _{4}\left( t\right)  &=&f_{F}^{+}\left( t,eV+\hbar \omega _{o}\right)
+f_{B}^{+}\left( t,-eV-\hbar \omega _{o}\right) \nonumber \\
&& +f_{F}^{-}\left( t,eV+\hbar
\omega _{o}\right) +f_{B}^{-}\left( t,-eV-\hbar \omega _{o}\right)   \nonumber
\\
&=&2\mathrm{Re}\left[ \xi _{1}^{s}(t)\right] ,  \label{coeff_electric4} \\
\Delta _{5}\left( t\right)  &=&f_{F}^{+}\left( t,eV-\hbar \omega _{o}\right)
+f_{B}^{+}\left( t,-eV+\hbar \omega _{o}\right) \nonumber \\
&& +f_{F}^{-}\left( t,eV-\hbar
\omega _{o}\right) +f_{B}^{-}\left( t,-eV+\hbar \omega _{o}\right)   \nonumber
\\
&=&2\mathrm{Re}\left[ \xi _{2}^{s}(t)\right] ,  \label{coeff_electric5}
\end{eqnarray}%
where $\xi _{1}^{s}(t)$ and $\xi _{2}^{s}(t)$ are defined in Eqs.~(\ref{xi1s}) and
(\ref{xi2s}), respectively.
By using Eqs.~(\ref{Markovian_limit})--(\ref{Markovian_real_B}) in the
Markovian WBL (i.e., taking $\Lambda_e \to\infty$), the above three terms
that contribute to the total current recover their respective Markovian
versions presented in Ref.~\onlinecite{Wabnig 2005}: 
\begin{eqnarray}
\frac{I_{\rm position}^{M}}{e} &=&eV\left[ G_{0}+2G_{x}\langle x(t)
+G_{xx}\langle x^{2}(t)\rangle %
\right], 
\label{Mar_x-current} \\
\frac{I_{P}^{M}}{e} &=&\frac{G_{p}\hbar }{m}\left\langle p\left( t\right)
\right\rangle ,  \label{Mar_p-current} \\
\frac{I_{QM}^{M}}{e} &=&-\frac{\hbar G_{xx}}{4m\omega _{o}}\left[ \left( eV+\hbar
\omega _{o}\right) \coth \frac{eV+\hbar \omega _{o}}{2k_{B}T}
\right. \nonumber \\
&&\qquad \qquad \left. -\left(
eV-\hbar \omega _{o}\right) \coth \frac{eV-\hbar \omega _{o}}{2k_{B}T}\right].
\label{Mar_QM-current}
\end{eqnarray}%

The last term in Eq.~(\ref{current}), $I_{\left\{ X,P\right\} }(t)/e $, is a new term that was not discussed in
Ref.~\onlinecite{Wabnig 2005}. 
This term originates from symmetrized product of the position and momentum
operators and can be written as 
\begin{equation}
\frac{I_{\left\{ X,P\right\} }(t)}{e}=\frac{iG_{xx}}{4m\omega _{o}\pi }%
\left[ \Delta _{6}\left( t\right) -\Delta \,_{7}\left( t\right) \right]
\left\langle \left( xp+px\right) \left( t\right) \right\rangle ,
\label{xp-current}
\end{equation}%
where%
\begin{eqnarray}
\Delta _{6}\left( t\right) &=&f_{F}^{+}\left( t,eV+\hbar \omega _{o}\right)
-f_{B}^{+}\left( t,-eV-\hbar \omega _{o}\right) \nonumber \\
&& -f_{F}^{-}\left( t,eV+\hbar
\omega _{o}\right) +f_{B}^{-}\left( t,-eV-\hbar \omega _{o}\right) \nonumber\\
&=&2i\mathrm{Im}\left[ \xi _{1}^{a}(t)\right] , \\
\Delta _{7}\left( t\right) &=&f_{F}^{+}\left( t,eV-\hbar \omega _{o}\right)
-f_{B}^{+}\left( t,-eV+\hbar \omega _{o}\right) \nonumber \\
&& -f_{F}^{-}\left( t,eV-\hbar
\omega _{o}\right) +f_{B}^{-}\left( t,-eV+\hbar \omega _{o}\right) \nonumber \\
&=&2i\mathrm{Im}\left[ \xi _{2}^{a}(t)\right] ,
\end{eqnarray}%
and $\xi _{1}^{a}(t)$ and $\xi _{2}^{a}(t)$ are defined in Eqs.~(\ref{xi1a}) and
(\ref{xi2a}), respectively.
In the Markovian limit, this term becomes
\begin{eqnarray}
\frac{I_{\left\{ X,P\right\} }^{M}}{e} &=&\frac{G_{xx}}{2m\omega _{o}\pi }\Bigl[\frac{%
{}}{{}}\Theta _{F}\left( eV+\hbar \omega _{o}\right) -\Theta _{B}\left(
-eV-\hbar \omega _{o}\right)   \nonumber \\
&&\qquad \qquad -\Theta _{F}\left( eV-\hbar \omega _{o}\right) +\Theta _{B}\left(
-eV+\hbar \omega _{o}\right) \Bigr]\nonumber \\
&&\qquad \quad \times \left\langle \left( xp+px\right) \left(
t\right) \right\rangle ,  \label{Mar_xp-current}
\end{eqnarray}
where the functions $\Theta _{F}(x)$ and $\Theta _{B}(x)$ are defined in
Eqs.~(\ref{Markovian_image_F}) and (\ref{Markovian_image_B}),
respectively. This extra current
term, Eq.~(\ref{xp-current}), with a coefficient coming from the
combination of the imaginary parts of the QPC reservoir correlation
functions, was generally ignored in the studies of the same problem in the
literature \cite{Mozyrsky 2002,Wabnig 2005,Clerk 2004,Smirnov 2003,Wabnig
2007}. The contribution to the diffusion coefficient $h_{e}(t)$, Eq.~(\ref%
{diffusion}), also comes from the imaginary parts of the QPC reservoir
correlation functions but with a different combination. Unlike the
diffusion coefficient $h_{e}(t)$ which is generally much smaller than the
other decoherence and damping coefficients for typical parameters and is
thus often neglected, we will show in the next
section that Eq.~(\ref{xp-current}) has a substantial contribution to the
total transient current in the non-Markovian case and differs
qualitatively and quantitatively from its Markovian WBL counterpart, Eq.~(\ref%
{Mar_xp-current}) with $\Lambda_e \to\infty$. Thus it may serve as a
witness or signature of finite-bandwidth
non-Markovian features for the coupled NMR-QPC system.

\section{Numerical Results and Analysis}
\label{sec:numerics} 

In our numerical calculations, we first concentrate on the case 
where the influence of the thermal environment is neglected. 
This allows us to address the
non-Markovian effect coming solely from the QPC reservoirs. This case where
the effect of the QPC reservoirs dominates over that of the thermal
environment may nevertheless be 
justified for a much larger relative coupling strength of QPC to the NMR
and for typical QPC bias voltages
and reservoir temperatures. 
We will discuss in subsection \ref{sec:thermal}
the case when the effect of the thermal environment is
included and is comparable to that of the QPC reservoirs.

\subsection{Effect of only the QPC reservoirs}

\begin{figure}[tb]
\includegraphics[width=\linewidth]{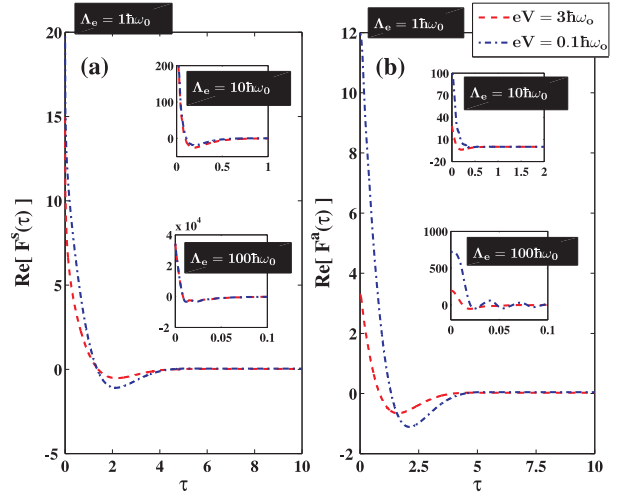}
\caption{(Color online) Real part of (a) the symmetric reservoir correlation
kernel (function) $F^{S}(\tau )$ and (b) the antisymmetric reservoir
correlation kernel (function) $F^{a}(\tau )$ at a small Lorentzian cut-off
energy of $\Lambda _{e}=\hbar \omega _{o}$ for different values of
the bias voltage: $(eV/\hbar \omega _{o})=0.1$ (blue dot-dashed lines), and $3$
(red dashed lines). The time is in units of $\omega_{o}^{-1}$. The insets in
(a) and (b) are for the cases of  $(\Lambda _{e}/\hbar \omega_{o})=10$ and $100$, respectively. Other parameter used is $k_{B}T=0.1\hbar\omega _{o}$.}
\label{fig:bath_CF}
\end{figure}

Figures \ref{fig:bath_CF}(a) and \ref{fig:bath_CF}(b) show the dependence of
real parts of the mode-independent two-time symmetric and antisymmetric QPC
reservoir correlation kernels (functions)  on the time-difference $\tau
=t-t_{1}$ for various values of the cut-off energy $\Lambda _{e}$. The
mode-independent two-time symmetric and antisymmetric QPC reservoir correlation
 kernels (functions) are evaluated by converting the summations over $%
k,k^{\prime }$ and $q,q^{\prime }$ of the mode-dependent correlation
functions (kernels) of Eqs.~(\ref{symmetrykernels}) and (\ref%
{asymmetrykernels}) into energy integrations with the Lorentzian spectral
density given by Eq.~(\ref{spectraldensity}), and then performing the energy
integrations. There are several characteristic time scales in this
non-Markovian problem. The time scale of the NMR is about $1/\omega _{o}$,
the time scale of the energy-dependent QPC spectral density is about $\hbar
/\Lambda _{e}$, the time scale of the applied bias
is $\hbar /eV$, the time scale of the QPC reservoir temperature is $\hbar
/k_{B}T$, the time scale of the electron tunneling is about 
$1/\Gamma_{AB}^{M}$ and the time scales of the combinations of the
electron tunneling rates are $p_0^2/D_{e}^{M}$, $1/\gamma_{e}^{M}$,
$\hbar/h_{e}^{M}$.  
We define the time scale at which the profiles of the QPC reservoir two-time
correlation kernels (functions) decay 
as the QPC bath correlation time, $\tau_{B}$. 
We can see from Figs.~\ref{fig:bath_CF}(a) and \ref{fig:bath_CF}(b) 
that at a given low cut-off energy of $\Lambda_e=\hbar\omega_o$ 
and at a low temperature of $k_B T=0.1\hbar \omega _{o}$, 
the reservoir two-time
correlation kernel (function) with a 
high bias voltage of $eV=3\hbar \omega _{o}$ (in red dashed lines)
has a slightly shorter reservoir correlation time $\tau _{B}$ than
that with a low
bias voltage of $eV=0.1\hbar \omega _{o}$ (in blue dot-dashed line). 
But as indicated from the insets of Figs.~\ref{fig:bath_CF}(a) and
\ref{fig:bath_CF}(b), the dependence of $\tau _{B}$
on the bias voltage is much weaker than that on the cut-off energy $\Lambda
_{e}$. The insets show that the larger the cut-off energy $\Lambda _{e}$ is,
the smaller the bath correlation time $\tau _{B}$ is. Moreover, the bath
correlation time in this case is about $\tau _{B}\sim \hbar /\Lambda _{e}$.

\begin{figure}[tb]
\includegraphics[width=\linewidth]{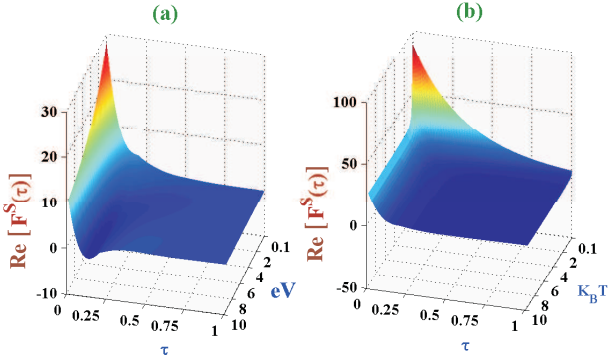}
\caption{(Color online) (a) Real part of the symmetric reservoir correlation
kernel (function) $F^{S}(\tau )$ as a function of the time-difference $\tau =t-t_{1}$ and the bias voltage $eV$. The bias
voltage is in units of $\hbar \protect\omega _{o}$ and the time is in units
of $\omega_{o}^{-1}$. Other parameters used are $\Lambda _{e}=\hbar 
\omega _{o}$ and $k_{B}T=0.1\hbar \omega _{o}$. (b) Real
part of the symmetric reservoir correlation kernel (function) $F^{S}(\tau)$
as a function of the time-difference $\tau =t-t_{1}$ and the QPC lead temperature $k_{B}T$. The lead temperatures is in units
of $\hbar\omega _{o}$ and the time is in units of $\omega_{o}^{-1}$. Other parameters used are $\Lambda _{e}=\hbar \omega _{o}$
and $eV=0.1\hbar \omega _{o}$.}
\label{fig:bath_CF2}
\end{figure}

Figure \ref{fig:bath_CF2}(a) shows the symmetric reservoir correlation
kernel (function) that depends on bias voltage $eV$ and time-difference $\tau
=t-t_{1}$ for a cut-off energy of $\Lambda _{e}=1.0\hbar \omega _{o}$.
Although the bath correlation time is affected mainly by the value of the
cut-off energy $\Lambda _{e}$, one can still see for the
Lorentzian spectral density Eq.~(\ref{spectraldensity}) chosen here
that when the bias voltage
decreases, the bath correlation time $\tau _{B}$ increases. Figure 
\ref{fig:bath_CF2}(b) shows the dependence of the symmetric reservoir
correlation kernel (function)
on temperature $k_BT$. Similar to that on the bias voltage, the
correlation time increases when the reservoir temperature decreases. 
The effect of the temperature on the bath correlation time seems to be
stronger than that of the bias voltage.
The Markovian approximation is valid in the case when the bath
correlation time $\tau _{B}$ is much smaller than the typical system
response time $\tau _{S}$.  The typical response time of our NMR
system is about the minimum value of  $(1/\omega _{o},1/\Gamma_{AB}
^{M},p_0^2/D_{e},1/\gamma _{e},\hbar/h_{e})$.

\begin{figure}[tb]
\includegraphics[width=\linewidth]{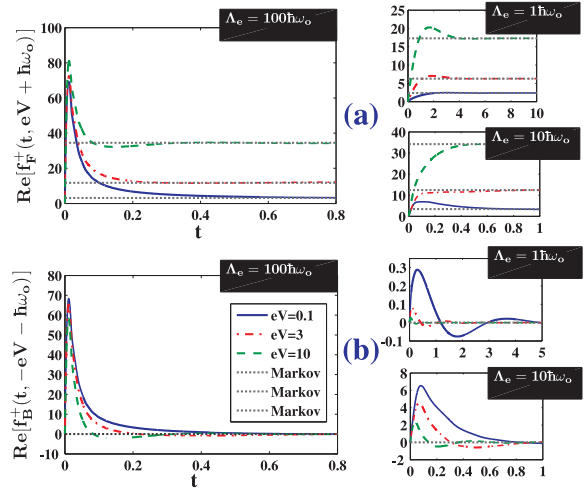}
\caption{(Color online) Real parts of the time-dependent forward
  (backward) inelastic 
electron tunneling coefficients for different values of the QPC bias
voltage: 
$(eV/\hbar \omega _{o})=0.1$ (in blue solid lines), $3$ (in red
dot-dashed lines), and $10$ (in green dashed lines). The Markovian cases
with finite cut-off energies are plotted in gray dotted lines. The time is
in units of $\omega_{o}^{-1}$. Other parameters used are $k_{B}T=0.1\hbar 
\omega _{o}$, $\Lambda _{e}=100\hbar \omega _{o}.$ The
small subplots in (a) and (b) are for the cases of $(\Lambda _{e}/\hbar \omega
_{o})=1$ and $10$, respectively.
Other parameter used is 
$k_{B}T=0.1\hbar \omega _{o}$.}
\label{fig:coeff_eVa}
\end{figure}

\begin{figure}[tb]
\includegraphics[width=\linewidth]{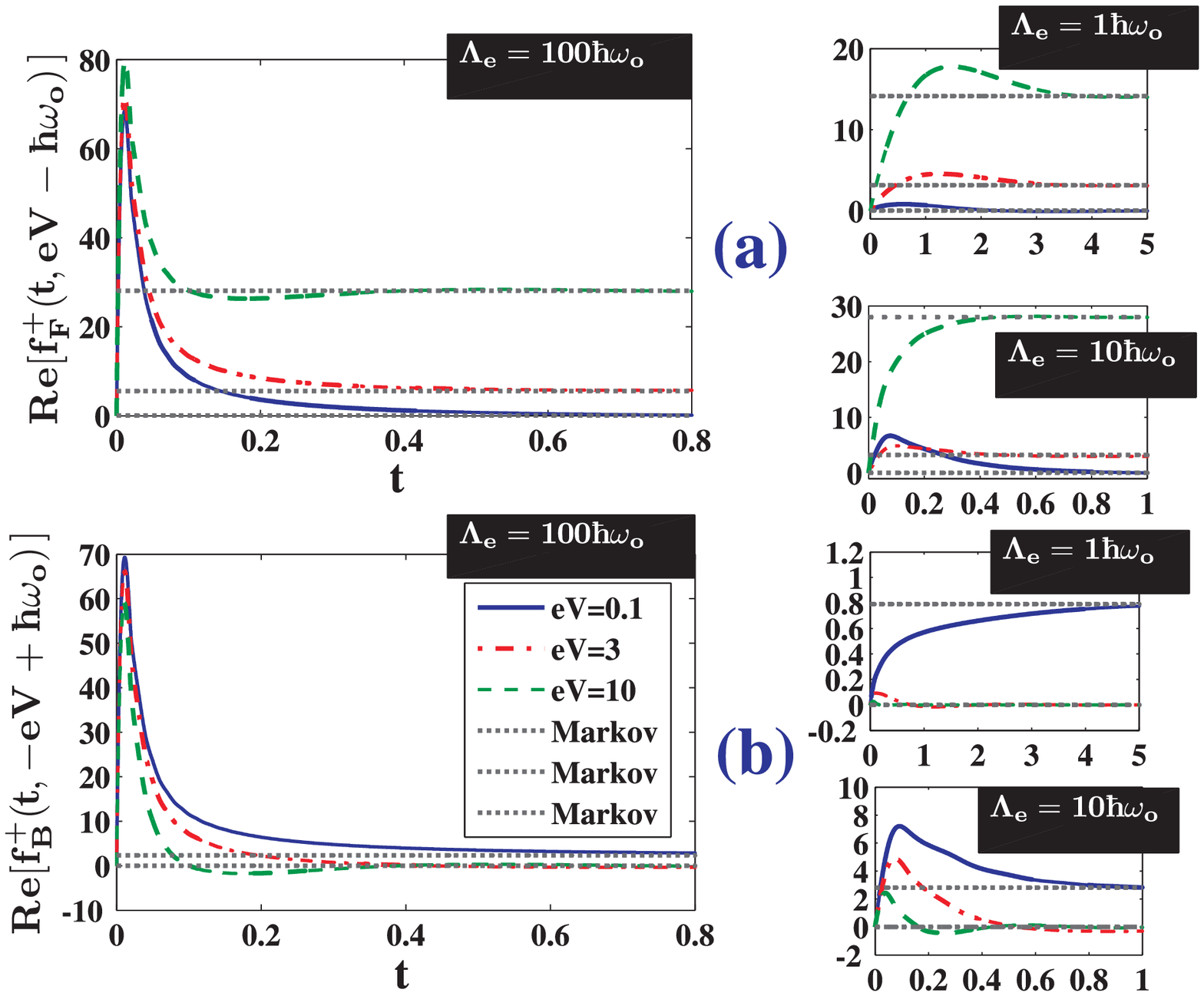}
\caption{(Color online) Real parts of the time-dependent forward
  (backward) inelastic 
electron tunneling coefficients for different values of the QPC bias
voltage: 
$(eV/\hbar \omega _{o})=0.1$ (in blue solid lines), $3$ (in red
dot-dashed lines), and $10$ (in green dashed lines). The Markovian cases
with finite cut-off energy are plotted in gray dotted lines. The time is
in units of $\omega_{o}^{-1}$. Other parameters used are $k_{B}T=0.1\hbar 
\omega _{o}$, $\Lambda _{e}=100\hbar \omega _{o}.$ The
small subplots in (a) and (b) are for the cases of $(\Lambda _{e}/\hbar \omega
_{o})=1$ and $10$, respectively.
Other parameter used is 
$k_{B}T=0.1\hbar \omega _{o}$.}
\label{fig:coeff_eVb}
\end{figure}

\begin{figure}[tb]
\includegraphics[width=\linewidth]{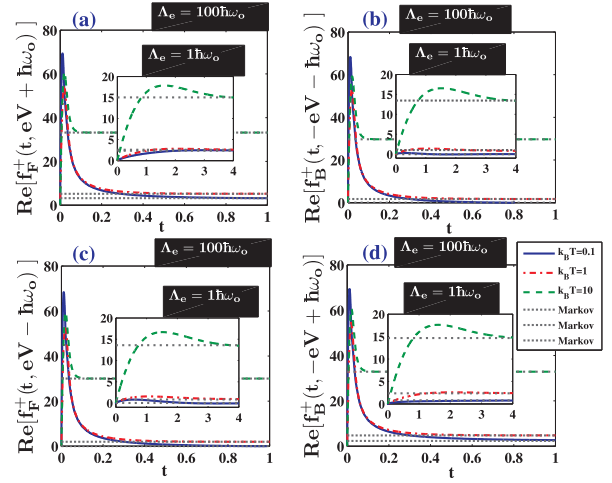}
\caption{(Color online)  Real parts of the time-dependent forward
  (backward) inelastic 
  electron tunneling 
coefficients for different values of the QPC lead temperatures: $(k_{B}T/\hbar 
\omega _{o})=0.1$ (in blue solid lines), $1$ (in red dot-dashed
lines), and $10$ (in green dashed lines). The Markovian cases with finite
cut-off energy are plotted in gray dotted lines. The time is in
units of $\omega_{o}^{-1}$. Other parameters used are $eV=0.1\hbar
\omega_{o}$, and $\Lambda _{e}=100\hbar \omega _{o}$. The inset in
each subplot is for the
case of $eV=0.1\hbar \omega _{o}$ and the small Lorentzian cut-off
energy of $\Lambda _{e}=\hbar\omega _{o}$. 
}
\label{fig:coeff_beta}
\end{figure}

Figures \ref{fig:coeff_eVa}, \ref{fig:coeff_eVb} and \ref{fig:coeff_beta} show
the real parts of some typical time-dependent coefficients 
$f_{F(B)}^{\pm}$ of the $n$-resolved master equation for different
values of the  bias voltage and temperature, respectively. 
The cut-off energy, $\Lambda _{e}$, is also varied in
each subplot. Physically, these coefficients, if multiplied by the factor $%
2g_{L}^{0}g_{R}^{0}A_{00}B_{00}/\hbar $ (where the value of $A_{00}$ and $B_{00}$ could
be either one of the tunneling amplitudes $T_{00}$ and $\tilde{\chi}_{00}$),
correspond to the time-dependent finite-temperature forward (backward)
inelastic QPC electron tunneling rates that accompany with the absorption or
emission of the NMR energy quanta. Compared to their Markovian counterparts
that are constants in time plotted in dashed lines in 
Figs. \ref{fig:coeff_eVa}, \ref{fig:coeff_eVb} and \ref{fig:coeff_beta}, 
the memory effects of the QPC reservoirs are contained in the time-dependent
coefficients. As the cut-off energy $\Lambda _{e}$ is increased, the
profiles (peaks)
of the time-dependent coefficients become higher and the half-maximum widths
become narrower. Moreover, the positions of the peaks of the profiles 
also shift to the
short time region. In other words, the memory effects of the non-Markovian
time-dependent coefficients persist for longer times for smaller values of
the cut-off energy. We distinguish the Markovian case, where the cut-off
energy $\Lambda _{e}$ is finite, from the Markovian WBL case, where the
spectral density becomes energy-independent as $\Lambda _{e}\rightarrow
\infty $. So one can also see that the time-dependent coefficients approach
their respective long-time Markovian counterparts of 
Eqs.~(\ref{Markovian_real_F}) and (\ref{Markovian_real_B}) with finite
cut-off energies (bandwidths).
Furthermore, the
non-Markovian coefficients with larger cut-off energies 
approach more rapidly in time to 
their Markovian counterparts. 

Generally speaking, for the inelastic forward tunneling coefficients
in Figs.~\ref{fig:coeff_eVa}(a), \ref{fig:coeff_eVb}(a),
\ref{fig:coeff_beta}(a) and \ref{fig:coeff_beta}(b),
when the values of the temperature and bias
voltage are higher, the long-time asymptotic (Markovian) values of the
time-dependent coefficients are larger but the time scales for the 
time-dependent coefficients approaching their long-time values become shorter.
The inelastic (emission) backward tunneling coefficients in 
Fig.~\ref{fig:coeff_eVa} (b) approach, at a rather low temperature of 
$k_{B}T=0.1\hbar \omega _{o}$, approximately zero (the Markovian value) at
long times since the argument of $(-eV-\hbar \omega _{o})$ is negative 
[see Eq.~(\ref{Markovian_real_B})]. Similarly, since the argument
of $(eV-\hbar \omega _{o})$ for a low bias voltage of 
$eV=0.1\hbar \omega_{o}$ is negative, 
the inelastic (emission) forward tunneling
coefficients in blue solid lines in Fig.~\ref{fig:coeff_eVb}(a) also
approach, at a low 
temperature, approximately zero (the Markovian value) at long
times. On the other hand in Fig.~\ref{fig:coeff_eVb}(b), since the
argument $(-eV+\hbar \omega _{o})$ for $eV=0.1\hbar \omega _{o}$ is
positive, the phonon-assisted backward tunneling 
is allowed even in the negatively biased direction at a low temperature and
thus the inelastic tunneling coefficients in blue solid lines become
finite at large times. One 
can see from Figs.~\ref{fig:coeff_eVa} and \ref{fig:coeff_eVb} that if the
bias voltage is increased, the inelastic forward tunneling coefficients in
Figs.~\ref{fig:coeff_eVa}(a) and \ref{fig:coeff_eVb}(a) are enhanced while
the inelastic backward tunneling coefficients in Figs.~\ref{fig:coeff_eVa}(b)
and \ref{fig:coeff_eVb}(b) are suppressed. 
Unlike the cases of $eV=0.1\hbar\omega$ in Figs.~\ref{fig:coeff_eVa}(b) and ~\ref{fig:coeff_eVb}%
(a), the time-dependent inelastic tunneling coefficients (in red
dot-dashed lines and in green dashed lines) in 
Figs.~\ref{fig:coeff_beta}(b) and \ref{fig:coeff_beta}(c) approach finite
nonzero values for large temperatures even though their arguments 
of $(-eV-\hbar \omega _{o})$ and $(eV-\hbar \omega _{o})$ are negative. 
Moreover, the non-Markovian
coefficients with a large cut-off energy of  $\Lambda_e=100\hbar\omega$ 
shown in Fig.~\ref{fig:coeff_beta} approach their corresponding
Markovian counterparts more rapidly as the value of the temperature becomes
higher. Similar to the case of increasing temperature, the time-dependent
non-Markovian coefficients with large cut-off energies saturate to their corresponding Markovian
counterparts more quickly as the bias voltage is increased as shown in 
Figs.~\ref{fig:coeff_eVa} and \ref{fig:coeff_eVb}.
In summary, the characteristic times  for the non-Markovian behaviors of the
time-dependent coefficients $f^{\pm}_{F(B)}$ to differ from their Markovian
counterparts are usually longer for smaller cut-off energies, bias voltages
and temperatures.

\begin{figure}[tb]
\includegraphics[width=0.92\linewidth]{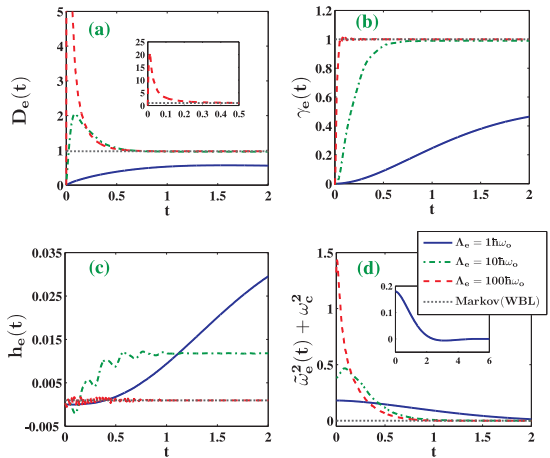}
\caption{(Color online) Time-dependent coefficients of the
  unconditional master equation: (a) decoherence
coefficient $D_{e}(t)$ in unit of $p_{o}^{2} \gamma_e^{M}$, (b)
damping coefficient $\gamma _{e}(t)$ in units of $\gamma_e^{M}$,
(c) diffusion coefficients $h_{e}(t)$ in units of $\hbar \gamma_e^{M}$
and (d) frequency renormalization shift $[\tilde{\omega}
_{e}^{2}(t)+\omega_c^{2}]$ in units of 
$\omega _{0} \gamma_{e}^M$ for different
values of the finite Lorentzian cut-off energy: 
$(\Lambda _{e}/\hbar\omega _{o})=1$ (in blue solid lines), $10$ (in
green dashed lines), and $100$
(in red dot-dashed lines). The Markovian WBL cases are plotted in gray
dotted lines. 
The time is in units of $\omega_{o}^{-1}$. 
The inset in (a) shows the large-value behavior of
$D_{e}(t)$ in the short-time regime for the case of 
$\Lambda _{e}=100\hbar\omega _{o}$, and the inset in (d) shows 
the long-time behavior of the frequency renormalization shift $[\tilde{\omega}
_{e}^{2}(t)+\omega_c^{2}]$ for the case of 
$\Lambda _{e}=\hbar\omega _{o}$. 
Other parameters used are $eV=0.1\hbar \omega _{o}$,
and $k_{B}T=0.1\hbar \omega _{o}$. 
}
\label{fig:coeff_uncond}
\end{figure}

Next we discuss the time-dependent coefficients of the unconditional
master equation defined in  
Eqs.~(\ref{renormalization})--(\ref{diffusion}).
We plot in Figs.~\ref{fig:coeff_uncond} (a), (b), (c) and (d)
the time-dependent decoherence coefficient $D_{e}(t)$, damping coefficient $%
\gamma _{e}(t)$, diffusion coefficients $h_{e}(t)$ and frequency
renormalization shift
$[\tilde{\omega}_{e}^{2}(t)+\omega_c^{2}]$
respectively, for
different values of the Lorentzian cut-off energy $\Lambda_e$. 
The time-dependent
coefficients are affected primarily by the values of cut-off energy $\Lambda
_{e}$ for small fixed values of bias voltage $eV=0.1\hbar \omega
_{o}$ and temperature $k_{B}T=0.1\hbar \omega _{o}$, and approach to their
long-time-limit values in a time scale of about $\hbar /\Lambda _{e}$. The
time-dependent coefficients with large cut-off energies saturate at their
corresponding Markovian WBL values (in dotted lines), while the coefficient
with a small cut-off energy of $\Lambda_{e}=\hbar \omega_o$
approach a long-time value different from the Markovian
WBL value. 
The contribution to the frequency normalization
$\tilde{\omega}_{e}^{2}(t)$ comes from the
imaginary part of the combination of the time-dependent tunneling
coefficients [see Eq.~(\ref{renormalization})] or the combination of
the QPC reservoir correlation kernels (functions). We can see 
from  Fig.~\ref{fig:coeff_uncond}(d) that the 
frequency renormalization shift
$[\tilde{\omega}_{e}^{2}(t)+\omega_c^{2}]$ approaches zero at large
times since the counterterm frequency contribution $\omega_c^{2}$
compensates the frequency renormalization at large times, 
i.e., $\omega_c^{2}=-(\tilde{\omega}_{e}^{M})^{2}$.
The diffusion coefficient $h_{e}(t)$ 
coming from the contributions of the imaginary part
of the combination of the tunneling coefficients 
[see Eq.~(\ref{diffusion})] or the imaginary part of the combination
of the reservoir correlation kernels or functions 
(or coming from the contributions of the Cauchy
principle values in the Markovian case) is typically very small compared to
the other coefficients (see Fig.~\ref{fig:coeff_uncond}) and thus is
often neglected in the discussion of the 
reduced dynamics of the NMR. We will show later that a transient current
term coming also from the contributions of the imaginary parts of the
reservoir kernels with a different combination has, however, a substantial
value and should be kept in order to describe correctly the measured
time-dependent current.

\begin{figure}[tb]
\includegraphics[width=\linewidth]{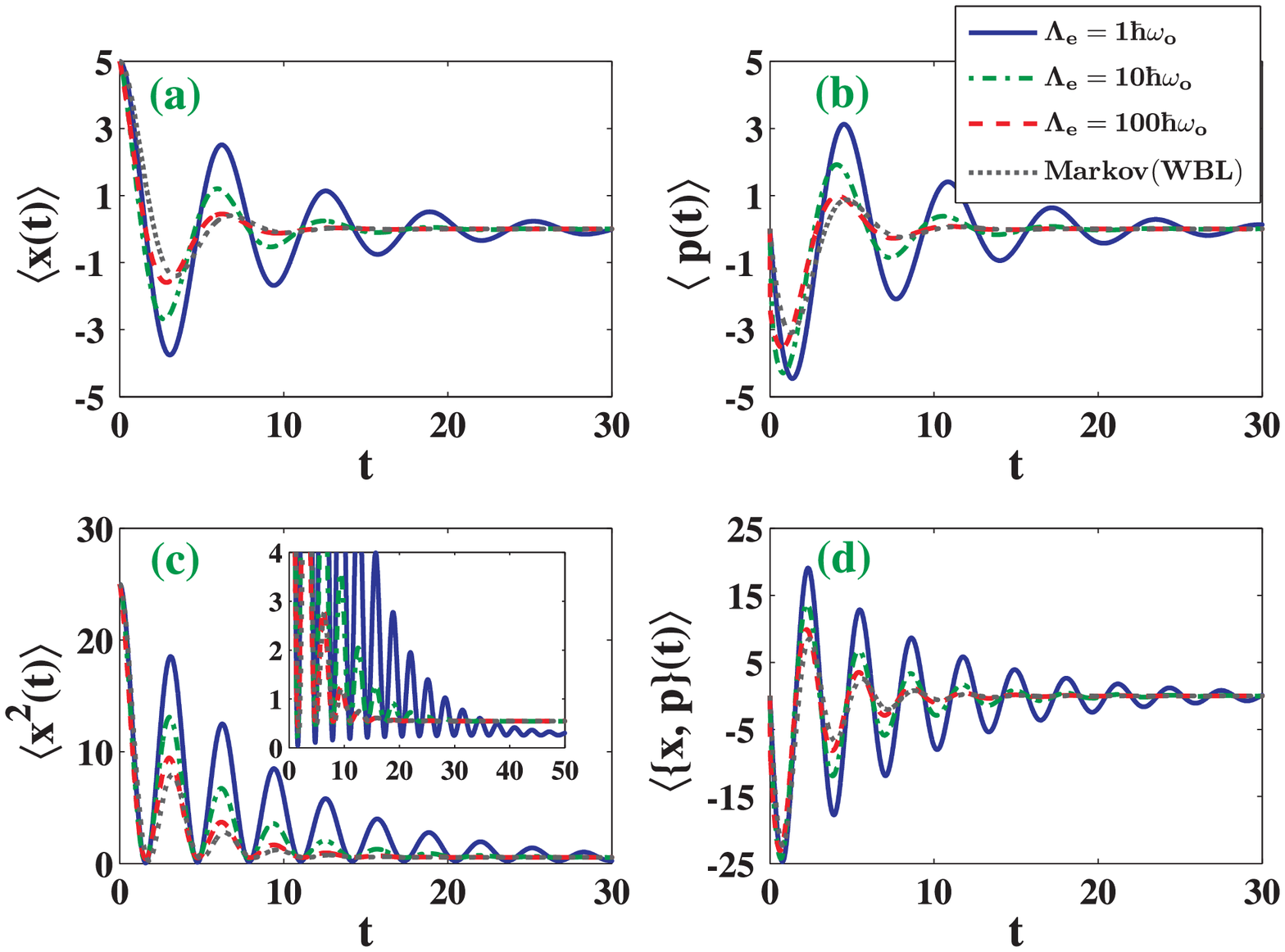}
\caption{(Color online) Time evolutions of the dynamical variables of the NMR for different values of the finite
Lorentzian cut-off energy: $(\Lambda _{e}/\hbar \protect\omega _{o})=1$ (in
blue solid lines), $10$ (in green dot-dashed lines), and $100$ (in red
dashed lines). (a)
position $\left\langle x(t)\right\rangle $ in units of $x_{0}$, (b) momentum $\left\langle
p(t)\right\rangle $ in units of $p_{0}$, (c) second moment position $\langle x^{2}(t)\rangle $ in units of $x_{0}^{2}$,
and (d) symmetrized second moment position-momentum 
$\langle \{x, p\}(t)\rangle$ in units of $x_{0}p_{0}$.  The Markovian
WBL cases are plotted in gray dotted lines. 
The NMR is initially in a coherent state with $\langle x(0) \rangle=5 x_0$
and  $\langle p(0) \rangle=0$.
The time is in units of $\omega_{o}^{-1}$.
The inset in (c) shows the long-time behavior of (c) for small
values of $\langle x^{2}(t)\rangle$.  
Other parameters used are $eV=0.1\hbar \omega _{o}$, 
$k_{B}T=0.1\hbar \omega _{o}$, 
and $\gamma _{e}^M=0.12\omega _{o}$.}
\label{fig:NMR_dynamics}
\end{figure}

Figure \ref{fig:NMR_dynamics} shows the numerical results of the
dimensionless mean and covariance values of the dynamical variables of the
NMR for different values of the cut-off energy. 
The dimensionless mean values $\langle x(t)\rangle /x_{0}
$ and $\langle p(t)\rangle /p_{0}$ oscillate with a frequency of about the
NMR renormalized frequency $\omega _{o}$ while the variances $\langle
x^{2}(t)\rangle /x_{0}^{2}$ and $\langle x(t)p\left( t\right) +p\left(
t\right) x(t)\rangle /x_{0}p_{0}$ oscillate with twice the frequency. As
expected, the non-Markovian results with larger cut-off energies $\Lambda
_{e}$ are closer to their Markovian WBL results, and there are considerable
differences between the Markovian WBL cases and the non-Markovian cases with
the cut-off energies $\Lambda _{e}$ comparable to the 
NMR frequency $\omega_{o}$.
We note here that our results include the frequency renormalization
in the non-Markovian cases. We can see from
Fig.~\ref{fig:coeff_uncond}(d) that the non-Markovian cases with larger
cut-off energies have a slightly larger renormalized physical
frequency in the short-time region. As a result, the initial
oscillatory behaviors 
of $\langle x(t)\rangle /x_{0}$ and $\langle p(t)\rangle /p_{0}$ 
with larger $\Lambda_{e}$ in Figs.~\ref{fig:NMR_dynamics}(a) and (b)
are slightly lower (i.e., with a slightly larger frequency) in the
short-time region than that of the Markovian WBL case (in dotted line).
Furthermore, one can also observe that the dynamical variables with larger
cut-off energies $\Lambda_{e}$ approach to their steady-state values
faster than those with smaller cut-off energies, and 
the steady-state values of the
dynamical variables 
in Fig.~\ref{fig:NMR_dynamics} match well the
values of the analytical expression of
Eqs.~(\ref{steady_state0})--(\ref{steady_state2}) and (\ref{stationaryxx}).

\begin{figure}[tb]
\includegraphics[width=0.9\linewidth]{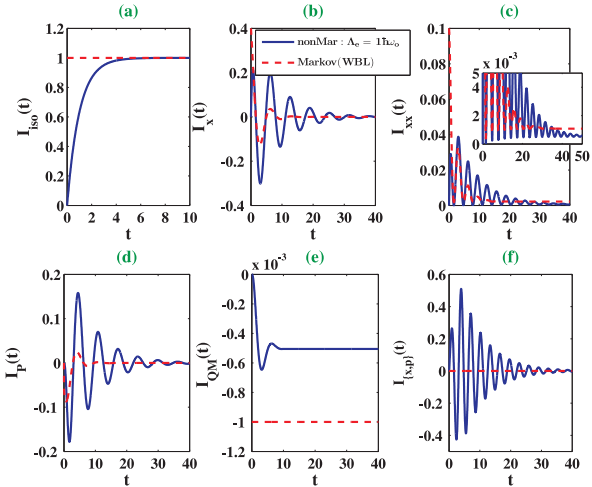}
\caption{(Color online) Individual contributions of the average
current $I(t)$ of Eq.~(\protect\ref{current}) for
Markovian WBL case (in red dashed lines) and the non-Markovian
cases with a small cutoff energy of $\Lambda _{e}=\hbar \omega _{o}$
(in blue solid lines). The first term in Eq.~(\protect\ref{current}), i.e.,
Eq.~(\ref{x-current}), is divided into three parts: (a) the isolated
QPC tunneling current, (b) the current proportional to $\langle x(t)\rangle $,
and (c) the current proportional to $\langle x^{2}(t)\rangle $. The last three
terms of the average current, Eq.~(\protect\ref{current}), are plotted
in (d), (e) and (f), respectively. Each individual contribution of the
average current is in units of $I^M_{\rm iso}=e^2VG_{0}$
and the time is in units of $\omega _{o}^{-1}$. 
The inset in (c) shows the long-time behavior of (c) for small
values of $I_{XX}(t)$.  
Other parameters used
are  $eV=0.1\hbar \omega _{o}$, $k_{B}T=0.1\hbar \omega _{o}$,
$G_{x}x_{0}/G_0=0.04$, $G_{p}x_{0}/G_0=0.01$ ,
$G_{xx}x_{0}^{2}/G_0=0.002$,  
and $\gamma _{e}^M=0.12\omega _{o}$.}
\label{fig:current}
\end{figure}

Figure \ref{fig:current} shows the differences in time evolutions of the
individual contribution terms of the average current, 
Eq.~(\ref{current}), between 
the Markovian WBL case and the non-Markovian cases with a small cut-off
energy of $\Lambda _{e}=\hbar \omega _{o}$. We further divide the first term
in Eq.~(\ref{current}), i.e., Eq.~(\ref{x-current}), into three parts: the
isolated QPC tunneling current, the current proportional to $\langle
x(t)\rangle $, and the current proportional to $\langle x^{2}(t)\rangle $. These
three parts are plotted in Figs.~\ref{fig:current}(a), (b) and (c),
respectively. The last three terms of the total current,
Eq.~(\ref{current}), are plotted in Figs.~\ref{fig:current}(d), (e) and
(f), respectively. 
There are considerable differences between the non-Markovian and Markovian
WBL results. In particular, the tunneling coefficients in the
Markovian WBL are constants in time, and thus the coefficients
$\Delta_i$ in front of the individual contributions to the total average 
current become also time-independent. As a result, the initial values of
the individual contribution terms of the average current will depend only
on the initial values of the dynamical variables of the NMR. 
For example, we choose an initial state such that $\langle x(0)\rangle\neq 0 $
 and $\langle x^{2}(0)\rangle\neq 0$. Then the Markovian WBL
current contributions $I_X(t)$ and $I_{XX}(t)$ start at a finite value
at time $t=0$, i.e., the QPC responses instantaneously to the motion of
the NMR to generate a finite current at $t=0$ (see the red dashed
lines in Fig.~\ref{fig:current}). 
This is of course not physical. 
In contrast, in the non-Markovian case, the tunneling
coefficients and  $\Delta_i(t)$ are time-dependent and their initial
values at the moment $t=0$ when the QPC detector is brought to
interact with the NMR are zeros. 
Thus the individual contribution terms
of the average current start from zeros 
(see the blue solid lines in Fig.~\ref{fig:current}) and will approach their
Markovian (finite-bandwidth) counterparts at a time scale of
$\hbar/\Lambda_e$.  
The extra transient current term $I_{\{X,P\}}(t)$ of
Eq.~(\ref{xp-current}), plotted in Fig.~\ref{fig:current}(f), is
proportional to the 
expectation value of the symmetrized product of the position and momentum
operators of the NMR $\langle\{x,p\}(t)\rangle$ and thus oscillates
with twice the frequency of $\omega_{o}$. 
This additional term, Eq.~(\ref{xp-current}), that was
generally ignored in the studies of the same problem in the literature 
\cite{Mozyrsky 2002,Wabnig 2005,Clerk 2004,Smirnov 2003,Wabnig 2007}
has a coefficient 
proportional to the imaginary part of the combination of the QPC
tunneling coefficients or bath
correlation kernels (functions) of 
$\mathrm{Im}\left[\xi_1^a- \xi _{2}^{a}\right]$,
where $\xi _{1}^{a}$ and $\xi _{2}^{a}$ are defined in Eqs.~(\ref{xi1a}) and
(\ref{xi2a}), respectively.
Recall that the frequency renormalization $\tilde{\omega}_{e}^{2}(t)$
of Eq.~(\ref{renormalization}) 
and the diffusion coefficient $h_e(t)$ of Eq.~(\ref{diffusion}) are
also proportional to the imaginary part of the QPC tunneling
coefficients or bath correlation kernels (functions) but with different
combinations.  
The value of the frequency normalization
$\tilde{\omega}_{e}^{2}(t)\propto \mathrm{Im}\left[\xi_1^a(t)+ \xi
  _{2}^{a}(t)\right]$ increases as $\Lambda_e$ increases, and diverges
as $\Lambda_e\to\infty$. So a counter term is introduced for the
purpose of frequency regularization. On the other hand, the value of
the diffusion
coefficient $h_e(t)\propto \mathrm{Im}\left[ \xi
_{1}^{s}\left( t\right) -\xi _{2}^{s}\left( t\right) \right]$
decreases as $\Lambda_e$ increases even though the individual terms of   
the imaginary parts of $\xi_{1}^{s}(t)$ and
$\xi_{2}^{s}(t)$ defined in Eqs.~(\ref{xi1s}) 
and (\ref{xi2s}) diverge as $\Lambda_e\to\infty$. 
The typical values of $h_e(t)$ are however very small as compared to
other time-dependent decoherence and dissipation coefficients 
(see Fig.~\ref{fig:coeff_uncond})
and thus $h_e(t)$ is often neglected.  
Similar to $h_e(t)$, the extra transient current 
$I_{\{X,P\}}(t)\propto [ \Delta_{6}( t) -\Delta_{7}( t)]\propto\mathrm{Im}\left[\xi_1^a(t)- \xi _{2}^{a}(t)\right]$
decreases as $\Lambda_e$ increases even though the individual terms of   
the imaginary parts of $\xi_{1}^{a}(t)$ and  $\xi_{2}^{a}(t)$ defined in Eqs.~(\ref{xi1a})
and (\ref{xi2a}) also diverge as $\Lambda_e\to\infty$. 
However, $I_{\{X,P\}}(t)$
has a considerable magnitude and should be included into the
time-dependent current. $I_{\{X,P\}}(t)$ is also proportional to 
$\langle\{x,p\}(t)\rangle$ which vanishes in the steady state, 
so $I_{\{X,P\}}(t)$ exits only in the transient regime.  
As mentioned that $I_{\{X,P\}}(t)$ decreases as $\Lambda_e$ increases,
the contribution of $I_{\{X,P\}}(t)$ to the transient current  
is very small for large cut-off energies. Indeed, we can see from
the red dashed line in Fig.~\ref{fig:current}(f) that $I_{\{X,P\}}(t)$
does not contribute to 
the transient current in the Markovian WBL case. 

\begin{figure}[tb]
\includegraphics[width=\linewidth]{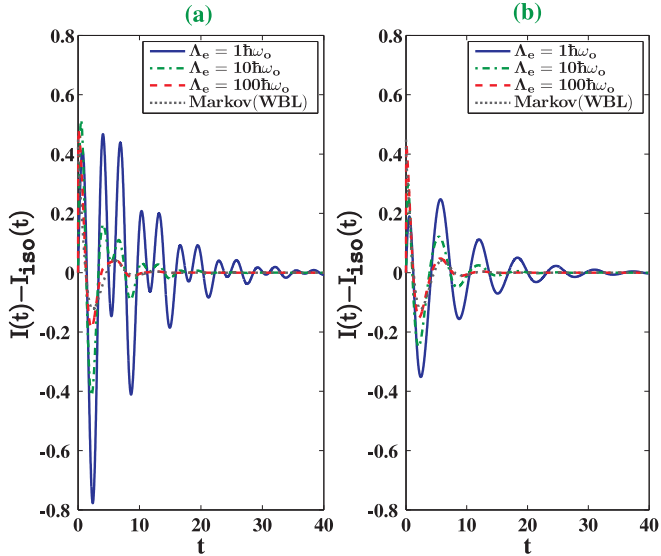}
\caption{(Color online) (a) Difference between the total current and the
isolated QPC tunneling current,
$I_{\mathrm{tot}}(t)-I_{\mathrm{iso}}(t)$, and
(b)current in (a) with the contribution of $I_{\{X,P\}}\left( t\right) $ of
Eq.~(\ref{xp-current}) furthermore deducted for different values of
the Lorentzian cut-off energy: $(\Lambda _{e}/\hbar \omega _{o})=1
$ (in blue solid lines), $10$ (in red dashed lines), and $100$ (in green
dot-dashed lines). The Markovian WBL cases are plotted in gray dotted lines.
The time-dependent average current is in units of $I^M_{\rm iso}=e^2VG_{0}$ and the
time is in units
of $\omega _{o}^{-1}$. Other parameters used are $eV=0.1\hbar \omega
_{o}$, $k_{B}T=0.1\hbar \omega _{o}$, 
$G_{x}x_{0}/G_0=0.04$, $G_{p}x_{0}/G_0=0.01$ ,
$G_{xx}x_{0}^{2}/G_0=0.002$, 
and $\gamma _{e}^M=0.12\omega _{o}$.}
\label{fig:current_diff}
\end{figure}

For clarity, we plot in Fig.~\ref{fig:current_diff}(a) the difference
between the total current and the isolated QPC tunneling current,
i.e., $I(t)-I_{\mathrm{iso}}(t)$, for different values of the cut-off
energy. For comparison, we further deduct the contribution of
$I_{\{X,P\}}(t)$
of Eq.~(\ref{xp-current}) from $I(t)-I_{\mathrm{iso}}(t)$, and
the resultant current is plotted in Fig.~\ref{fig:current_diff}(b). 
In the steady state, the expectation values 
$\left\langle x\right\rangle _{t\rightarrow \infty } =\left\langle
p\right\rangle _{t\rightarrow \infty }=\left\langle \left\{
  x,p\right\} \right\rangle _{t\rightarrow \infty }=0$. 
Thus the total steady-state
average current that approaches its Markovian
long-time value becomes    
\begin{equation}
I(t\to\infty)=I_{\rm iso}(t\to\infty)+I_{XX}(t\to\infty)
+I_{QM}(t\to\infty).
\end{equation}
It has been discussed in 
Ref.~\onlinecite{Wabnig 2005} that in the limit of small bias voltages and
temperatures (i.e., $eV,k_BT \ll \hbar\omega_o$), the quantum
correction current ${I_{QM}}(t\to\infty)$ in the steady state 
has to cancel $I_{XX}(t\to\infty)$
as the voltages and temperatures are too small to excite the NMR. 
As a result, the steady-state average
current in this case is equal to that of an isolated QPC junction, 
$I_{\rm iso}(t\to\infty)$.
This is indeed the case 
for the low values of voltage and temperature chosen in
Figs.~\ref{fig:current} and \ref{fig:current_diff}.
We can see from Fig.~\ref{fig:current}(e) and the inset of 
Fig.~\ref{fig:current}(c) that the steady-state ${I_{QM}}(t\to\infty)$ does
cancel the steady-state $I_{XX}(t\to\infty)$. As a result, 
the total average current difference $I(t)-I_{\mathrm{iso}}(t)$ 
in Figs.~\ref{fig:current_diff} vanishes in the steady state. 
We can also see from Figs.~\ref{fig:current_diff}(a) and (b) that for a large
cut-off energy or bandwidth of $\Lambda_e=100\hbar\omega_0$, the
evolution of the time-dependent current approaches that of
the Markovian WBL case closely.  
Without including the contribution
of $I_{\{X,P\}}(t)$ of Eq.~(\ref{xp-current}) to the average current,
 there are still substantially
quantitative difference between the Markovian WBL current and the
non-Markovian currents with finite cut-off energies as shown in
Fig.~\ref{fig:current_diff}(b).  Furthermore, both
significantly qualitative and quantitative differences between the
non-Markovian currents and the Markovian WBL current in the short-time
region are clearly 
observed in Fig.\ref{fig:current_diff}(a). The non-Markovian transient
currents with small values of cut-off energy are characterized by
oscillations with large amplitudes and twice the NMR renormalized frequency as
compared to the Markovian WBL one. 
We note that although 
a significant
difference between the non-Markovian and 
Markovian WBL currents  $I_{XX}(t)$ in Fig.~\ref{fig:current}(c) can
be observed, the values of $I_{XX}(t)$ are small compared to those of
the other individual current contribution terms. So this difference
in $I_{XX}(t)$ may not be easily identified in the total average
current as that of $I_{\{X,P\}}(t)$.    
This extra contribution of $I_{\{X,P\}}(t)$ of
Eq.~(\ref{xp-current}) was completely neglected in the discussion of the
Markovian current for the same problem in the literature \cite{Mozyrsky
2002,Wabnig 2005,Clerk 2004,Smirnov 2003,Wabnig 2007}. We find, however,
that this extra significant contribution of $I_{\{X,P\}}(t)$ in the transient
current may serve as a witness or signature of the non-Markovian features
for the coupled NMR-QPC system with finite cut-off energies (bandwidths).

\subsection{Inclusion of the effect of thermal bosonic environment}
\label{sec:thermal}

\begin{figure}[tb]
\includegraphics[width=\linewidth]{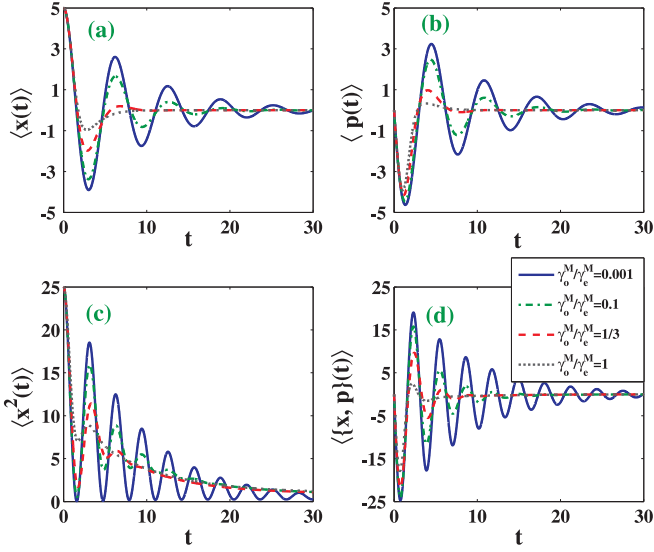}
\caption{(Color online) Time evolutions of the dynamical variables
  of the NMR (a)
position $\left\langle x(t)\right\rangle $ in units of $x_{0}$, (b) momentum $\left\langle
p(t)\right\rangle $ in units of $p_{0}$, (c) second moment position $\langle x^{2}(t)\rangle $ in units of $x_{0}^{2}$,
and (d) symmetrized second moment position-momentum 
$\langle \{x, p\}(t)\rangle $ in units of $x_{0}p_{0}$ for different values of 
the ratio of 
$\gamma_o^M/\gamma_e^M=0.001$ (in
blue solid lines), $0.1$ (in green dot-dashed lines), and $1/3$ (in
red dashed lines), and $1$ (in gray dotted line). 
The NMR is initially in a coherent state with $\langle x(0) \rangle=5 x_0$
and  $\langle p(0) \rangle=0$.
The time is in units of $\omega_{o}^{-1}$.
Other parameters used are $eV=0.1\hbar \omega _{o}$, 
$k_{B}T=0.1\hbar \omega _{o}$, $\Lambda_e=\Lambda_0=\hbar\omega_o$,
and $\gamma_{e}^M=0.12\omega _{o}$.}
\label{fig:dynamics_thermal}
\end{figure}

\begin{figure}[tb]
\includegraphics[width=\linewidth]{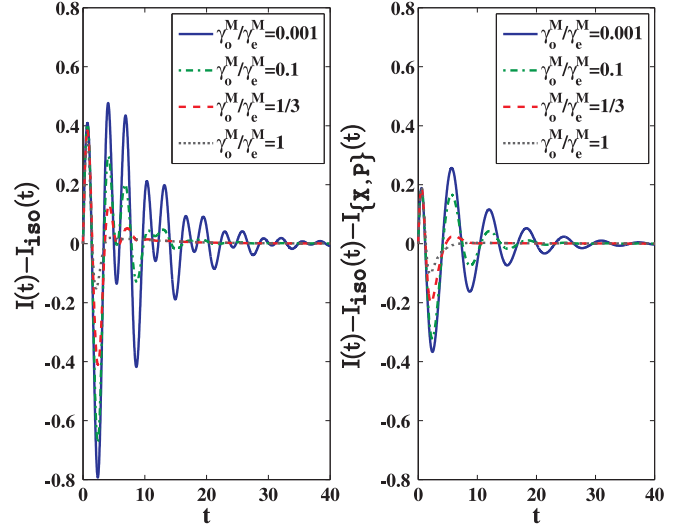}
\caption{(Color online) (a) Difference between the total current and the
isolated QPC tunneling current,
$I_{\mathrm{tot}}(t)-I_{\mathrm{iso}}(t)$, and
(b) current in (a) with the contribution of $I_{\{X,P\}}\left( t\right) $ of
Eq.~(\ref{xp-current}) furthermore deducted for different values of
the ratio of 
$\gamma_o^M/\gamma_e^M=0.001$ (in
blue solid lines), $0.1$ (in green dot-dashed lines), and $1/3$ (in
red dashed lines). and $1$ (in gray dotted line). 
The time-dependent average current is in units of $I^M_{\rm iso}=e^2VG_{0}$ and the
time is  in units of $\omega _{o}^{-1}$. 
Other parameters used are $eV=0.1\hbar \omega
_{o}$, $k_{B}T=0.1\hbar \omega _{o}$, $\Lambda_e=\Lambda_0=\hbar\omega_o$,
$G_{x}x_{0}/G_0=0.04$, $G_{p}x_{0}/G_0=0.01$ ,
$G_{xx}x_{0}^{2}/G_0=0.002$, 
and $\gamma _{e}^M=0.12\omega _{o}$.}
\label{fig:current_thermal}
\end{figure}

So far, we do not include the effect of the thermal bosonic environment
into our numerical calculations. 
A natural question is whether including the 
effect of the bosonic environment changes the 
picture that the transient current can be used to witness 
non-Markovian effects of the coupled NMR-QPC system. 
Notice that most of the individual time-dependent current terms depend on the
product of two factors: the combination of time-dependent coefficients
$\Delta_i(t)$, and the time-dependent dynamical variables of the NMR.  
The inclusion of the effect of the thermal bosonic environment affects
only the time evolution of the NMR dynamical variables but not the
time-dependent coefficients $\Delta_i(t)$. 
So if the coupling of the NMR to the thermal
environment is small compared with the coupling to the QPC reservoirs,
then the main non-Markovian feature in the QPC transient current 
will remain. But if the coupling of the NMR to the thermal
environment is comparable to the coupling to the QPC reservoirs, then
the dynamical variables will reach their steady-state more quickly. 
Figure \ref{fig:dynamics_thermal} shows the time evolutions of 
$\langle x(t)\rangle$, $\langle p(t)\rangle$, $\langle x^2(t)\rangle$,   
and $\langle\{x,p\}(t)\rangle$ for different values of the ratio of
$(\gamma_o^M/\gamma_e^M)$ that characterizes the coupling
strength of the NMR-thermal bath relative to that of the NMR-QPC reservoirs.  
One can see in Fig.~\ref{fig:dynamics_thermal}
that as the value of the ratio $(\gamma_o^M/\gamma_e^M)$
increases for a fixed value of $\gamma_e^M=0.12\omega_o$,
the oscillation amplitudes of $\langle x(t)\rangle$,
$\langle p(t)\rangle$, $\langle x^2(t)\rangle$, and
$\langle\{x,p\}(t)\rangle$ diminish and the time at which the
steady state is reached shifts to the short time region since the
total damping coefficient become larger. 
As a result, the differences in oscillation amplitudes between
$I(t)-I_{\mathrm{iso}}(t)$ and $I(t)-I_{\mathrm{iso}}(t)- I_{\{X,P\}}(t)$
become small and the time intervals where the differences exist with
characteristic oscillation frequency of $2\omega_o$ also become
shorter(see Fig.~\ref{fig:current_thermal}).   
The oscillations will quickly reach their steady-state values and 
differences will become unobservable if the ratio of $(\gamma_o^M/\gamma_e^M)$
becomes much larger than one.

\begin{figure}[tb]
\includegraphics[width=\linewidth]{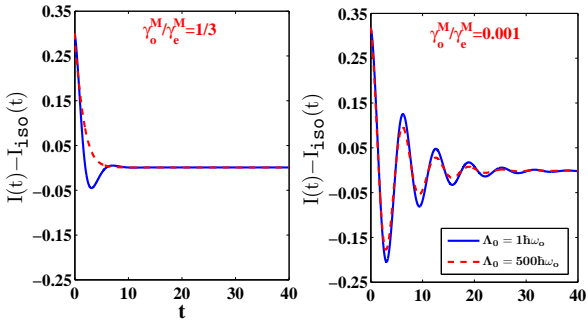}
\caption{(Color online)  
Current difference of $I(t)-I_{\mathrm{iso}}(t)$,
for different values of
the ratio of 
(a) $\gamma_o^M/\gamma_e^M=1/3$ and (b)  $\gamma_o^M/\gamma_e^M=0.001$.
The QPC reservoirs are in the WBL (i.e., $\Lambda_e\to\infty$), while
the frequency
bandwidth of the spectral density of the thermal bosonic bath 
is $\Lambda_0=\omega_0$ (in blue solid lines) and
$\Lambda_0=500\omega_0$ (in red dashed lines).
The time-dependent average current is in units of $I^M_{\rm iso}=e^2VG_{0}$ and the
time is in units of $\omega _{o}^{-1}$. 
Other parameters used are $eV=0.1\hbar \omega
_{o}$, $k_{B}T=0.1\hbar \omega _{o}$,
$G_{x}x_{0}/G_0=0.04$, $G_{p}x_{0}/G_0=0.01$ ,
$G_{xx}x_{0}^{2}/G_0=0.002$, 
and $\gamma _{e}^M=0.12\omega _{o}$.}
\label{fig:QPC_Mar_current}
\end{figure}

Another question is whether the case 
where there are no non-Markovian effects in the NMR-QPC system, but 
there are non-Markovian effects induced by the bosonic environment, 
will result in similar non-Markovian features in the transient current.
The answer to the question 
can be found as follows. 
In our simple Lorentzian spectral
density model, no non-Markovian effects in the NMR-QPC system implies that
the cut-off energy or bandwidth of the QPC reservoir 
spectral density is very large 
(i.e., $\Lambda_e\gg\hbar \omega_o$ or in the WBL) 
since the Markovian limit can be
justified by this condition.
In the Markovian case, all the coefficients of $\Delta_i(t)$ become
time-independent.  
As mentioned previously, the time-dependent coefficient of  
$[ \Delta_{6}-\Delta_{7}]\propto\mathrm{Im}[\xi_1^a- \xi
_{2}^{a}]$ of the extra transient current $I_{\{X,P\}}(t)$
decreases as $\Lambda_e$ increases and becomes very small in the
(Markovian) WBL.
So $I_{\{X,P\}}(t)$ is very small in the Markovian WBL  
[see, e.g., Fig.~\ref{fig:current}(f)] even though
$\langle\{x,p\}(t)\rangle$ has a considerable amplitude in the
transient regime [see, e.g., Fig.~\ref{fig:NMR_dynamics}(d)].
Therefore, if there are no non-Markovian effects in the
NMR-QPC system (i.e., in the WBL), the extra transient 
current term $I_{\{x,p\}}(t)$ does not contribute
even though there may be still significant 
oscillation amplitudes in $\langle\{x,p\}(t)\rangle$.
Figure \ref{fig:QPC_Mar_current} shows the current difference of
$I(t)-I_{\mathrm{iso}}(t)$ for different values of
the ratio of $(\gamma_o^M/\gamma_e^M)$ in the case where the QPC reservoirs
are in the Markovian WBL (i.e., $\Lambda_e \to \infty$). 
The non-Markovian feature of 
oscillations with frequency of $2\omega_o$ in 
$I(t)-I_{\mathrm{iso}}(t)$ is unobservable in 
Fig.~\ref{fig:QPC_Mar_current} as $I_{\{X,P\}}(t)$ does
not contribute and $I_{XX}(t)$ is too small.
The slight differences in $I(t)-I_{\mathrm{iso}}(t)$ between the case of
$\Lambda_o=1$ and $\Lambda_o=500$ 
are primarily due to the differences in 
$\langle x(t)\rangle$ and $\langle p(t)\rangle$
induced by the non-Markovian bosonic environment
for the two different values of the spectral density frequency
bandwidth $\Lambda_o$.

\section{Conclusions}
\label{sec:conclusion}

In summary, we have derived second-order time-local (time-convolutionless)
non-Markovian conditional ({\em n-}resolved) and unconditional master
equations of the reduced density matrix of a NMR subject to a measurement by
a low-transparency QPC or tunnel junction detector and an influence by a
thermal environment. Our non-Markovian master equations implemented with the
reservoir memory correlation prescription going beyond the WBL 
allow us to study the memory
effect of the non-equilibrium QPC fermionic reservoir and the equilibrium
bosonic thermal bath on the NMR. Our non-Markovian master equations
with time-dependent coefficients reduce,
in appropriate limits, to various Markovian versions of master equations in
the literature. Furthermore, our non-Markovian master equations are valid
for arbitrary temperatures of the 
thermal environment and QPC reservoirs (detector), and for
arbitrary bias voltages as long as the perturbation theory up
to the second order in the system-detector and system-environment coupling
strength holds.

We have found considerable differences in dynamics between the non-Markovian
cases and their Markovian counterparts. The fact that the QPC detector
induces a back action on the NMR and the motion of the NMR modulates the
current through the QPC are taken into account self-consistently. We have
also calculated the time-dependent transport current through the QPC which
contains information about the measured NMR system. We have found an extra
transient current term of $I_{\{X,P\}}(t)$ of Eq.~(\ref{xp-current}). This
extra term, with a coefficient coming from the combination of the imaginary
parts of the QPC reservoir correlation functions, was generally ignored in
the studies of the same problem in the literature. But we find that it has a
substantial contribution to the total transient current in the non-Markovian
finite-bandwidth case and differs qualitatively and quantitatively
from its Markovian WBL 
counterpart. Thus it may serve as a witness or signature of non-Markovian
features for the coupled NMR-QPC system.

\begin{acknowledgments}
We would like to acknowledge support from the National Science Council,
Taiwan, under Grant No. 97-2112-M-002-012-MY3, support from the Frontier and
Innovative Research Program of the National Taiwan University under Grants
No. 97R0066-65 and No. 97R0066-67, and support from the focus group program
of the National Center for Theoretical Sciences, Taiwan. We are grateful to
the National Center for High-performance Computing, Taiwan, for computer
time and facilities.
\end{acknowledgments}

\appendix





\end{document}